\documentclass[preprint]{aastex}

\usepackage{epsfig}

\newcommand{\sgmemail}{mashnik@t2y.lanl.gov}

\def\lsim{\lower.5ex\hbox{$\; \buildrel < \over \sim \;$}}
\def\gsim{\lower.5ex\hbox{$\; \buildrel > \over \sim \;$}}

\begin{document}

\title{
{\it Los Alamos National Laboratory Report LA-UR-00-3658, Los Alamos (2000)}\\
}

\title{
On Solar System and Cosmic Rays Nucleosynthesis and Spallation Processes
}

\author{Stepan G. Mashnik}
\affil{
T-16, Theoretical Division,
Los Alamos National Laboratory\\
Los Alamos, NM 87545\\
\sgmemail}

\begin{abstract}
%\noindent ~~~~~~~~~~~~~~~~~~~~~~~~~~~~~~~~~~~~~~~~~~~~~~~~ABSTRACT

\noindent 
A brief survey of nuclide
abundances in the solar-system and in cosmic rays
and of the believed mechanisms of 
their synthesis is given. 
The role of spallation processes in nucleosynthesis is discussed.
A short review of recent measurements, compilations, calculations,
and evaluations of spallation cross sections relevant to nuclear
astrophysics is given as well.
It is shown that in some past astrophysical simulations,
old experimental nuclear data and theoretical
cross sections 
that are 
%not compatible 
in poor agreement
with recent
measurements and calculations
were used.
New astrophysical simulations using recently measured 
and reliably calculated nuclear cross sections,
further researches in obtaining better cross sections,
and production of evaluated spallation cross sections
libraries for astrophysics            
are suggested.
\end{abstract}


\section{Introduction}
A considerable success was
achieved over the last decades  
in determination of abundances of nuclides in
the solar system and in cosmic rays as well as in understanding the
mechanisms of their synthesis (see, e.g., 
%\cite{abible,40years,crosas96,mcwilliam97,ramaty98,kappeler98,cameron99,
%busso99,kappeler99}
Burbidge, Burbidge, Fowler, \& Hoyle 1957; 
Crosas \& Weisheit 1996;
McWilliam 1997;
Wallerstein et al. 1997;
Ramaty et al. 1998;
K\"appeler, Thielemann, \& Wiescher 1998;
Cameron 1999;
Bethe 1999;
Busso, Gallino, \& Wasserburg 1999;
Ginzburg 1999;
Hamann \& Ferland 1999;
Henley \& Schiffer 1999;
K\"appeler 1999;
Khlopov 1999;
Salpeter 1999;
Wolfenstein 1999).
Nevertheless, many interesting questions still remain. So,
the light-element abundances, especially that of beryllium, and
the origin of low-energy cosmic rays and their role in the light-element
production require a critical reexamination 
(Ramaty et al. 1998).
%\cite{r}. 
Another open
question on chemical evolution in galaxies is, e. g., the fact that plots of
abundances relative to hydrogen,
[Be/H] and [B/H] versus [Fe/H] in halo stars both exhibit a slope of 
+1, rather than the value +2 that is expected for normal supernova
recycling of interstellar material 
(Crosas \& Weisheit 1996; Duncan et al. 1992).
%\cite{crosas96,duncan92}.

\newpage

Another open question is related with the effect of hypothetical
sources of non-equilibrium particles on the radiation-dominant (RD)
stage of expanding hot Universe, like the effect of antiproton interaction
with $^4$He on abundance of light elements (Khokhlov 1999). The abundance of
light elements is much more sensitive to possible effects of non-equilibrium 
particles than to the spectrum of the thermal electromagnetic background, so
more complete analysis of effects of non-equilibrium  particles on the RD
stage of the Universe is still to be performed in the future (Khokhlov 1999).

Note that in spite of a determinative role of nuclear astrophysics
in understanding mechanisms of nucleosynthesis and of a great
improvement of nuclear data over the past decades,
some of the remaining questions about abundances of both stellar
and inerstellar elements are related with uncertainties of
nuclear data used in astrophysics. So, one of the biggest remaining 
uncertainties in nuclear astrophysics today concerns the precise 
parameters of a pair of resonance levels in $^{16}$O, just below the
thermonuclear energy range (Salpeter 1999). 
%\cite{salpeter99}
Also, some of the important
for astrophysics nuclear reactions are either not measured yet, or the
results of recent measurements and model calculations 
by nuclear physicists
are
little known and not widely used yet by astrophysicists.

At the same time, some questions about elemental abundances, especially
of the interstellar light elements, are related more with the 
cosmology itself
and with elementary particle physics
% and their interconection
(Khokhlov 1999; Salpeter 1999; Turner \& Tyson 1999) 
%\cite{turnertyson99}
rather than with the ``old" nuclear
physics. So, as mentioned by Salpeter (1999),
%\cite{salpeter99}
to predict today's interstellar abundances quantitatively we need to know
how many stars of various masses were born and have already died, since only
in old age (e.g., planetary nebulae) and death (supernovae) does
the material from a star's interior reach interstellar space. This mass
distribution,
the ``initial mass function," is still somewhat uncertain
(Salpeter 1999).
%\cite{salpeter99}

The aim of the present paper is to review briefly the believed
today 
mechanisms of nucleosynthesis and the elemental abundances of stellar
and iterstellar matter and to highlight places where nuclear
spallation processes are important.
Nuclear spallation is our field of research for decades, so 
we hope to find points where
our experience and knowledge may help to a little better 
understanding some astrophysical questions.

\section{The
Solar System 
and
Cosmic Rays
Abundances 
%and Nucleosynthesis 
of 
Elements}

%In this section, we like to 
Let us
briefly review in the beginning 
the believed today
scenario of the origin of elements and of their abundances, so that
we may discuss later a possible contribution
to nucleosynthesis
from
spallation processes. 

The abundance of the solar system
elements is shown in Fig. 1. Data shown by the thick black
curve are taken from
Table 38  by Lang (1980)
and are based upon measurements of Type I carbonaceous chrondite meteorites
(meteorites containing carbon compounds with a minimum of stony
or metallic chrondite metals, and are thought to be of a better representation
than 
the old Suess and Urey's (1956) curve (thin, blue) 
%\cite{suessurey}
which was based on measurements of
terrestrial, meteoric, and solar abundances. 
%\cite{lang}.

%\eject

\begin{figure}[h!]
\vspace*{+1.0cm}
\hspace*{5mm}
\psfig{figure=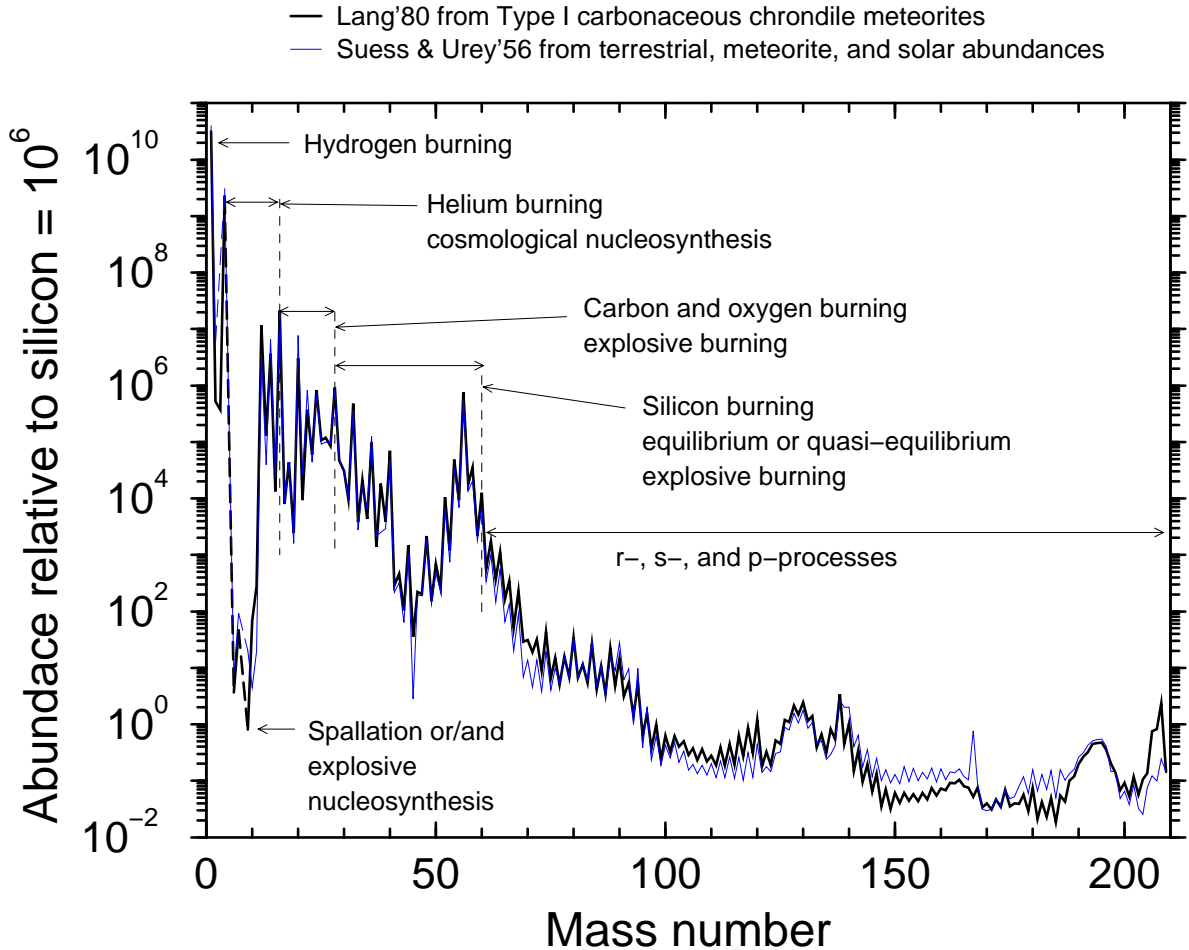,width=120mm,angle=270}
\vspace*{0.5cm}

\caption{
Abundances of solar system nuclides plotted as a function of mass number.
The thin blue curves shows old data compiled in Table III
by Suess and Urey (1956) which are based on measurements of terrestrial,
meteoric, and solar abundances. These data were used by Burbidge,
Burbidge, Fowler, and Hoyle (1957) in postulating the basic nucleosynthetic 
processes in stars in their seminal work which become widely known
as ``B$^2$FH," the ``bible" of nuclear astrophysics. 
The thick black curve shows newer data from the compilation
published in Table 38 by Lang (1980)
which are based upon measurement of Type I carbonaceous chrondite meteorites,
and are thought to be a better representation than Suess and Urey's curve.
The nuclear processes which are thought to be the main stelar mechanisms
of nuclide production are shown as well in the figure.
}
\end{figure}

An example of abundances of several light 
and medium elements in 70-280 MeV/nucleon 
cosmic rays
compared with the corresponding
abundances of elements in the solar system 
is shown in Fig. 2. One can see that while 
abundances 
of the majority of elements
in cosmic rays are very close to what we have
for the solar system, there are groups of nuclides, like the one
in the Sc-V-Mn region and, especially, the light LiBeB group, 
whose abundances in cosmic rays 
are
many orders of magnitude lower
than in the solar system.

%\eject

\begin{figure}[h!]
%\vspace*{+1.0cm}
\hspace*{5mm}
\psfig{figure=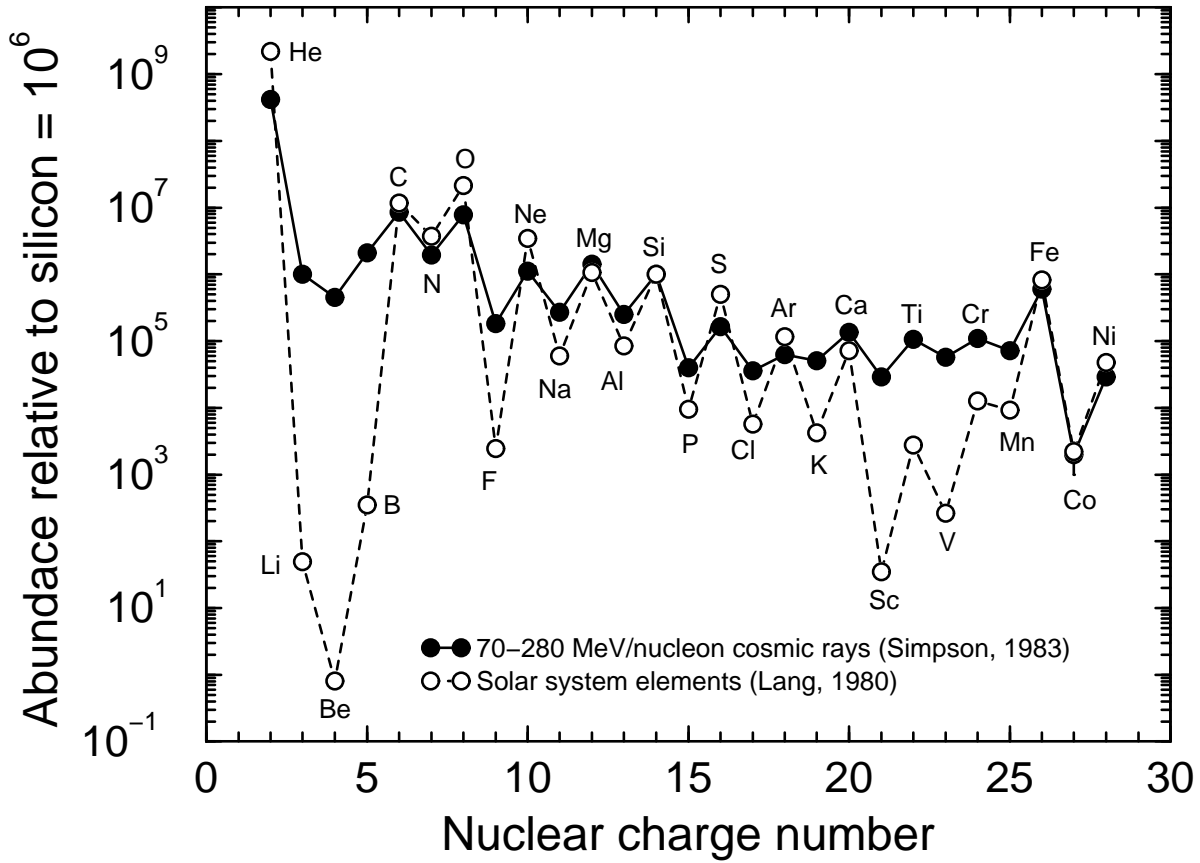,width=120mm,angle=270}
\vspace*{0.5cm}

\caption{
The relative elemental abundances of 70-280 Mev/nucleon cosmic rays 
(closed circles, taken from Tab.~2 by Simpson, 1983)
compared to the solar system abundances
(open circles, taken from Tab.~38 by Lang, 1980) normalized to 
Si = 10$^6$.
}
\end{figure}

\vspace*{0.5cm}
It is natural that abundances shown in both Figs. 1 and 2
are not definitive. 
With development of better measurement methods and techniques
and with increasing our general understanding of the astrophysics,
more reliable data will be obtained in the future. 
As one can see from
Table 1 (adopted from Schramm, 1995), not only the
precision of measurements increases with time but even the
objects of observation of elements and their presumed origins
change considerably in the course of time. Nevertheless,
the two sets of data shown in
Fig. 1 suggest us that one may expect no sweeping changes 
%with time 
in the main features of already measured
abundances of the solar system elements. So, even not definitive, these
abundances can be used confidently to study and to
understand the origin of elements.

%\eject

\begin{table*} [h]
\caption{Twenty-five Years for the Light Elements
(adopted from Schramm, 1995)}
\label{tab1}

\vspace*{3mm}
\begin{center}\begin{tabular}{l|ll||ll}
%\hline
& \multicolumn{2}{|c||}{25 YEARS AGO} & \multicolumn{2}{c}{PRESENT}   \\ 
\hline
\hline
Isotope & Best Observation & Presumed Origin & 
Best Observation & Presumed Origin \\
\hline
    D & Sea Water & T-Tauri Stars & ISM (HST) & BBN\\ 
$^3$He&Solar Flares &Low Mass Stars & Galactic H-II Regions & BBN plus\\ 
      &             &               & planetary Nebulae     & Low Mass Stars\\ 
$^4$He&Indirect     &BBN            & Extragalactic         & BBN     \\ 
      &             &               & H-II Regions          &          \\ 
$^7$Li&Pop I Stars  &T-Tauri  Stars & Pop II Stars          & BBN (Pop I has\\ 
      &             &               &                & Additional Sources\\
$^6$Li&Meteorites   &T-Tauri  Stars & Pop II Stars          & 
Cosmic Ray Spallation\\ 
Be&Pop I Stars   &T-Tauri  Stars & Pop II Stars          & 
Cosmic Ray Spallation\\ 
B&Meteorites   &T-Tauri  Stars & Pop II Stars (HST)    & 
Cosmic Ray Spallation\\ 
&&&& (Possible Additional\\ 
&&&& Source for $^{11}$B\\ 
 \end{tabular} \\[0.5ex]
\end{center} 
\end{table*}

It is believed today that the elements we
observe at present have been
generated mainly by three different processes (Reeves 1994):
The first one is the primordial nucleosynthesis, i.e., 
via thermonuclear reactions in 
the first few minutes
after the Big Bang and prior to formation of stars (this concerns mainly
D, $^3$H, $^3$He, $^4$He, $^7$Li, and perhaps some of the observed 
today Be and B; heavier elements could be produced by 
primordial nucleosynthesis, but were probably
burned thereafter in nuclear reactions during 
the stellar era). The second mechanism generating most of the observed
nuclei is 
nucleosynthesis in stars
(most of elements heavier than Li).
A third contribution to nucleosynthesis
comes from spallation reactions
in the interstellar medium (a part of the observed Li, Be, B, and some
heavier nuclides).
By convention, into the last group of nuclide production mechanisms 
can be included as well nuclear reactions induced by $\nu$
(see, e.g., Ryan et al. 1999; Khokhlov 1999), although
$\nu$-process nucleosynthesis is considered every so often in the
literature as a special mechanism (Woosley et al. 1990).
Let us discuss briefly below all these processes in turn.

\section{Big-Bang Nucleosynthesis (BBN)}

According to modern concepts, at time 
$t \simeq 15$ s after the Big Bang the
temperature of the Universe would have been 
%equal 
decreased
to $T \simeq 3 \times 10^9$ K 
and nucleosynthesis would then  begin through the
synthesis of deuterium from protons
\begin{equation}
\mbox{p} + \mbox{p} \to \mbox{D} + e^+ + \nu_e .
\end{equation}
This would have been the end of the ``radiative era,"
when radiation existed separately from matter as hadrons and leptons,
and the beginning of the ``nucleosynthesis era".

Note that the binding energy of the nucleons in deuterium is very small,
of only 2.2 MeV, which corresponds to
$T \sim 2.5 \times 10^{10}$ K. 
Therefore, at this stage,
almost all deuterium produced 
is rapidly destroyed 
by high-energy photons
and further synthesis of heavier
nuclei 
by means of reactions
\begin{equation}
\mbox{D} + \mbox{D} \to ^3\mbox{H} + \mbox{p} ,
\end{equation}
\begin{equation}
^3\mbox{H} + \mbox{D} \to \mbox{n} + ^4\mbox{He} ,
\end{equation}
\begin{equation}
n + ^3\mbox{He} \to ^3\mbox{H} + \mbox{p} ,
\end{equation}
is not possible until the temperature of the Universe decreases
to a value of 
$T \sim 10^{9}$ K. 
With further decrease in temperature the photodisintegration of 
deuterons practically ceases and deuterons begin to accumulate.
At the same time almost all of the neutrons are utilized in 
the creation of helium through the reaction (4).
By this time neutron decay would have shifted the neutron-proton balance to
13\% of neutrons and 87\% of protons
(see, e.g., Fig. 3.13 by Tsipenyuk 1997). 
This moment of time corresponds approximately to the third minute
after the Big Bang and to a temperature of $\sim 10^9$ K.

Beside reactions (1-4), there are other ways to get $^3$He
and $^4$He from nucleons during the BBN. So, the following reactions
are usually considered along with (1-4) to produce Helium from
Hydrogen at the BBN stage:
\begin{equation}
\mbox{p} + \mbox{n} \to \mbox{D} + \gamma ,
\end{equation}
\begin{equation}
\mbox{p} + \mbox{D} \to ^3\mbox{He} + \gamma,
\end{equation}
\begin{equation}
\mbox{n} + ^3\mbox{He} \to ^4\mbox{He} + \gamma ,
\end{equation}
\begin{equation}
\mbox{D} + ^3\mbox{He} \to ^4\mbox{He} + \mbox{p} ,
\end{equation}
\begin{equation}
\mbox{n} + \mbox{D} \to ^3\mbox{H} + \gamma ,
\end{equation}
\begin{equation}
\mbox{p} + ^3\mbox{H} \to ^4\mbox{He} + \gamma ,
\end{equation}
\begin{equation}
\mbox{D} + \mbox{D} \to ^3\mbox{He} + \mbox{n} .
\end{equation}

Nuclei which are heavier than helium would not have been produced 
in significant quantities during this time interval as there are no stable 
nuclei in the Nature with the mass numbers 5~and~8. Therefore two energy 
gaps would have appeared and synthesis of heavier nuclei would
have stopped for some time. The gap at $A = 5$ is overcome and the 
production of $^7$Li, $^7$Be, and $^6$Li together with their subsequent
possible destruction proceed through:
\begin{equation}
^3\mbox{H} + ^4\mbox{He} \to ^7\mbox{Li} + \gamma , 
\end{equation}
\begin{equation}
^7\mbox{Li} + \mbox{p} \to ^4\mbox{He} + ^4\mbox{He} , 
\end{equation}
\begin{equation}
^7\mbox{Li} + \mbox{D} \to ^4\mbox{He} + ^4\mbox{He} + \mbox{n} , 
\end{equation}
\begin{equation}
^3\mbox{He} + ^4\mbox{He} \to ^7\mbox{Be} + \gamma , 
\end{equation}
\begin{equation}
^7\mbox{Be} + \mbox{n} \to ^7\mbox{Li} + \mbox{p} , 
\end{equation}
\begin{equation}
^7\mbox{Be} +\mbox{D} \to ^4\mbox{He} + ^4\mbox{He} + \mbox{p} , 
\end{equation}
\begin{equation}
^4\rm{He} + \rm{D} \to ^6\rm{Li} + \gamma , 
\end{equation}
\begin{equation}
^6\rm{Li} + \rm{p} \to ^7\rm{Be} + \gamma , 
\end{equation}
\begin{equation}
^6\rm{Li} + \rm{n} \to ^7\rm{Li} + \gamma . 
\end{equation}

The gap at $A = 8$ prevents primeval production of heavier isotopes 
in any significant quantities. 
Generally, it should be mentioned that
different models assume different numbers of chains considered in BBN
calculations. So, while some 
authors limit themselves to only 12 most important reactions
in their BBN calculations
(see, e.g., Smith et al. 1993;
Sarkar 1999), other recent works consider up to 22 possible chains
(Lopez \& Turner 1998), or even much more extended nuclear networks
(see, e.g., Thomas et al. 1993) like the one  
shown in Fig. 3, kindly supplied by Keith Olive. 

Usually, one tries to determine
the primeval ratios of abundances of 
%primordial 
nuclei produced before
the ``star era" began, avoiding in observations regions where
the remnant matter from the Big Bang was processed through stars.
So, although all stars start on the main sequence and produce light
elements in their interiors, it is believed that most of the
observed today  interstellar helium was already there when the galaxy 
was formed, i.e., most of it is primordial and not from stars
(Salpeter 1999).
One reason for this is that there is little mixing from a star's 
center to its surface (and usually little mixing between stars and 
interstellar gas); another reason is that much of the interior helium 
is processed into heavier elements before a star dies.

The primordial
abundances of $^4$He, D, $^3$He, $^7$Li, and other light elements
measured in such a way are used further to fit the main parameters
of the BBN. The ``standard model" of the big bang nucleosynthesis,
in which it is assumed that the baryon distribution was uniform and
homogeneous during that period, is described by only one parameter, $\eta$,
the baryon to photon ratio, or by the baryon density, 
$\rho _B$, 
related to $\eta$ by 
$\rho _B = 6.88 \eta \times 10^{-22}$ g cm$^{-3}$ (Burles et al. 1999).
In practice, usually the baryon density is expressed not directly 
units of
$\rho _B$ but by a related parameter
$\Omega _B h^2$, where 
$\Omega _B$ is the baryon density in terms of the critical
mass density, $\rho _c$: 
$\Omega _B = \rho _B / \rho _c$, where 
$\rho _c = 1.88 \times 10^{-29} h^2$ g cm$^{-3}$ and $h$ is related to
the Hubble constant, $H_0$, by the relation
$H_0 = 100 h$ km s$^{-1}$ Mpc$^{-1}$ 
(see, e.g., Lang 1999).
%\eject
\newpage

\begin{figure}[h!]
\vspace*{-0.1cm}
\hspace*{5mm}
\psfig{figure=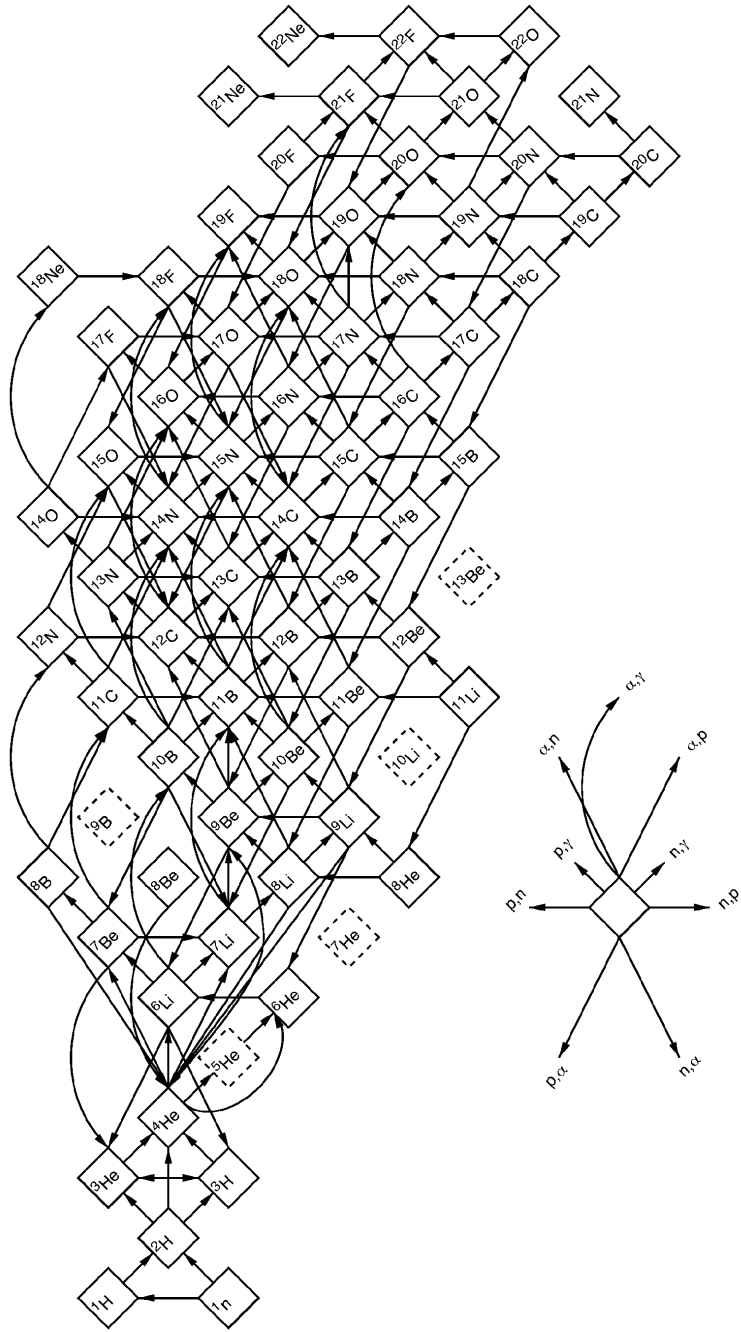,width=135mm,angle=0}
%\vspace*{0.5cm}

\caption{
The nuclear chain considered in the largest network of 
the Homogeneous BBN
calculations by Thomas, Schramm, Olive, and Fields (1993),
with kind permission from
Keith Olive. 
}
\end{figure}

\eject

As one can see from Fig. 4 (Turner 1999), a reasonable agreement 
between predicted
by the BBN abundances of 
$^4$He, D, $^3$He, $^7$Li 
and recent measurements may be achieved only in a narrow range of
values for the baryon density, namely,
$\Omega_Bh^2 = 0.019 \pm 0.0024$.

When we go
beyond the Standard Model, there is another fundamental parameter
which affects the BBN abundances, namely, the number of massless neutrino 
species in the Universe, $N_{\nu}$, which affects the expansion 
temperature-time relation and hence the way in which nuclear reactions
go out of thermal equilibrium. The presence of additional neutrino flavors
(or of any other relativistic species) at the time of nucleosynthesis
increases the energy density of the Universe and hence the expansion
rate, leading to a larger value of the temperature at the 
freeze-out of the weak-interaction rates, $T_f$, to a larger value
of $\rm{n}/\rm{p}$ ratio, and 
ultimately, to a higher value of the primeval $^4$He
abundance, $Y_{\rm{p}} = 2(\rm{n}/\rm{p}) / [1 + \rm{n}/\rm{p})]$. 
By means of a likelihood
analysis on $\eta$ and $N_{\nu}$ based on $^4$He and $^7$Li it was found 
that the 95\% CL range are  $1.7 \le N_{\nu} \le 4.3$ (Cassco 1998).
As one can see from Fig. 5, adapted from Copi, Schramm, \& Turner (1997), 
a recent analysis  
of the deuterium abundance in high-redishift hydrogen clouds helps to
sharpen this limit to $N_{\nu} \le 3.4(3.2)$, for $Y_{\rm{p}} = 0.242$, and
to $N_{\nu} \le 3.8(4.0)$, for $Y_{\rm{p}} = 0.252$, that is in a good
agreement with the Standard Model's value of
$N_{\nu} = 3$.
This fact can be treated as one more confirmation of the dramatic success of
the Big Bang model, which provides agreement with the observed element 
abundances only if the number of massless neutrino species is three, 
which correspond exactly to the three species (electron, muon,
and tau) we know to exist.

Besides the mentioned above two fundamental parameters, the BBN calculations
involve also a number of ``working" parameters, namely, the cross sections
for processes considered in the BBN nuclear networks. More exactly,
traditionally, in astrophysics are used not directly nuclear cross
sections but the so called ``nuclear reaction rates" derived from
measured or evaluated cross sections of relevant reactions convoluted with a
thermal (Maxwell-Boltzmann) relative velosity distribution. Useful references
on reaction rates works performed 
before 1993 can be found in Smith, Kawano,
\& Malaney (1993). The last and most complete compilation of reaction rates
involving light $(1 \le Z \le 14)$, mostly stable, nuclei,
called NACRE (Nuclear Astrophysics Compilation of REaction rates), 
have been published
recently by a big consortium of nuclear physics and astrophysical
European laboratories (Angulo et all. 1999), where further detailed
references may be found
(see the recent work by Vangioni-Flam, Coc, Casse, \& Oberto (2000),
where NACRE have been already used in an updated BBN model to study
primordial abundances of light elements up to $^{11}$B). 

When we have already  fixed
the two fundamental parameters of the BBN, $\eta$
and $N_\nu$ and have chosen the ``working horses", the needed 
thermonuclear reaction rates, we can perform BBN calculations to
study how abundances of different light elements have changed with the
time (or temperature) after the big bang, like shown in Fig. 6,
adapted from 
Burles, Nollett, \& Turner (1999).

\begin{figure}
\centerline{\psfig{figure=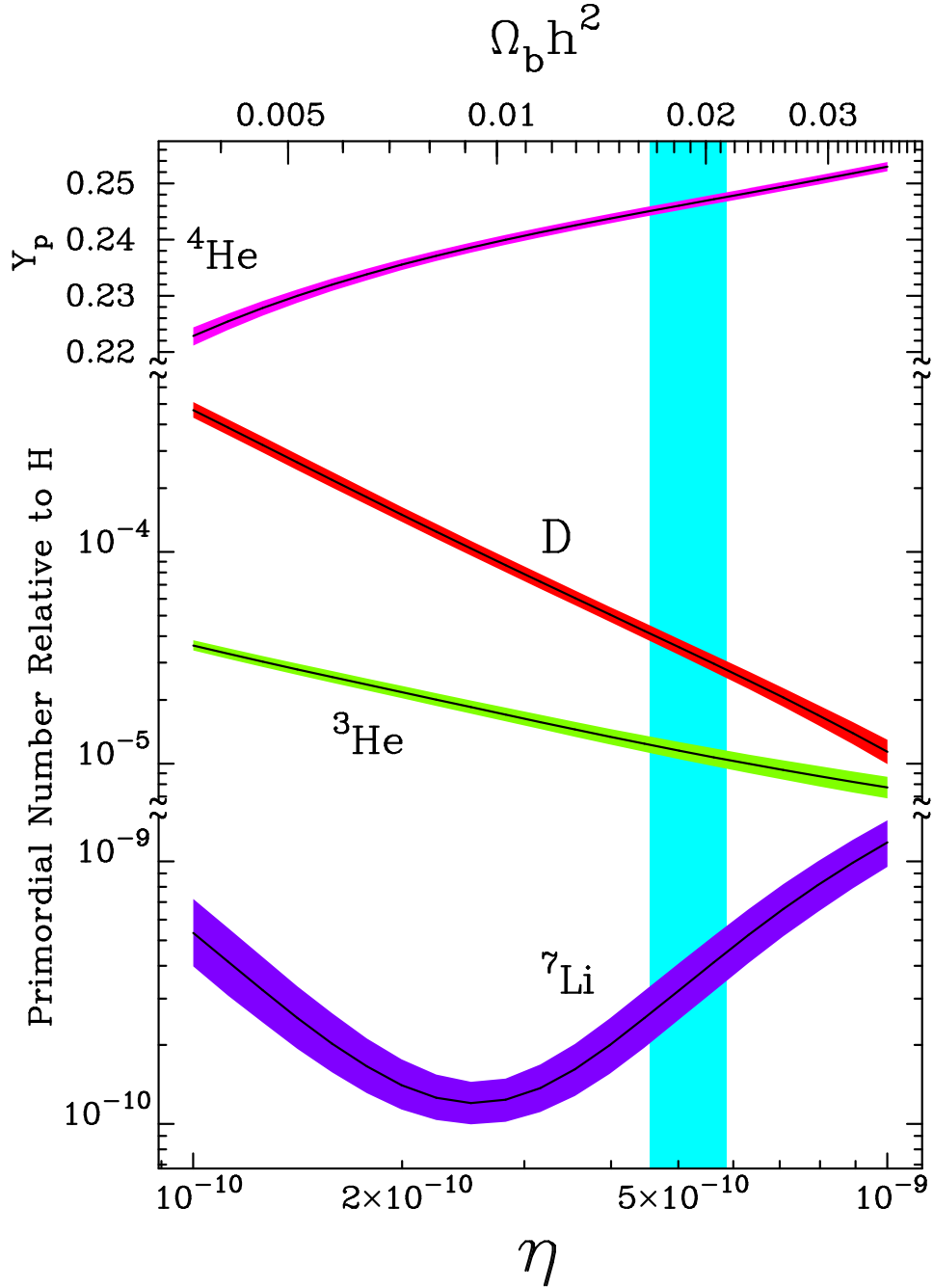,width=5in}}
\caption{Predicted abundances of $^4$He (mass fraction),
D, $^3$He, and $^7$Li (number relative to hydrogen) as a function
of the baryon density; widths of the curves indicate
``$2\sigma$'' theoretical uncertainty.  The dark band highlights
the determination of the baryon density based upon
the recent measurement of the primordial abundance of
deuterium (Burles \& Tytler, 1998a,b), $\Omega_Bh^2 = 0.019 \pm
0.0024$ (95\% cl); the baryon density is related to the baryon-to-photon
ratio by $\rho_B = 6.88\eta \times 10^{-22}\,{\rm g\,cm^{-3}}$
(Burles et al, 1999).
[From Turner 1999, 
%courtesy 
with kind permission from
Michael Turner].
}
\label{fig:bbn}
\end{figure}

%\eject

\begin{figure} 
\hspace*{5mm}
\center
\psfig{figure=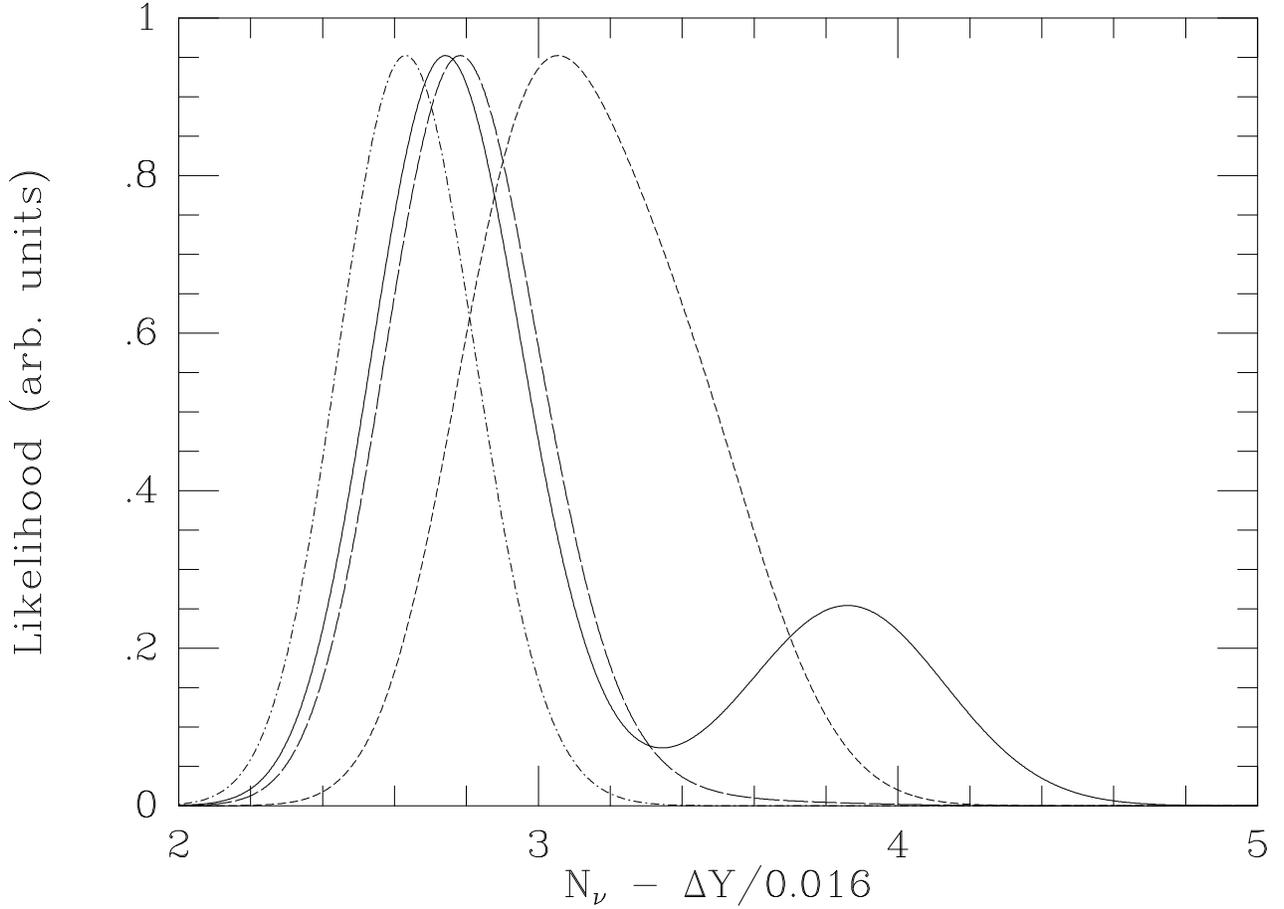,width=120mm,angle=270}
\vspace*{0.5cm}
\caption{Marginal likelihood for ${\tilde N} \equiv
N_\nu -\Delta Y/0.016$ with different Bayesian priors for the
primeval deuterium abundance:
(D/H)$_P \le 1.0$ (solid line); [(D + $^3$He)/H]$_P \le 2\times 10^{-4}$
(short-dashed line); extreme model of $^3$He chemical evolution
(from 
%Ref.~\protect\cite{ttsc}
Turner et al. 1996) (long-dashed line);
(D/H)$_P = (2.5\pm 0.5)\times 10^{-5}$ (dashed-dotted line).
In each case we have assumed the $^7$Li abundance that results
in the least stringent limit to $\tilde N$.  The fact that
$\tilde N = 3$ is well within the 95\% credibility interval
is indicative of the consistency of big-bang nucleosynthesis
with three massless neutrino species.
[From Copi, Schramm, and Turner, Phys. Rev. {\bf C55},
3389 (1997), 
%courtesy 
with kind permission from
Michael Turner.]}
\end{figure}

%\newpage

\begin{figure} 
%\hspace*{5mm}
\center
\psfig{figure=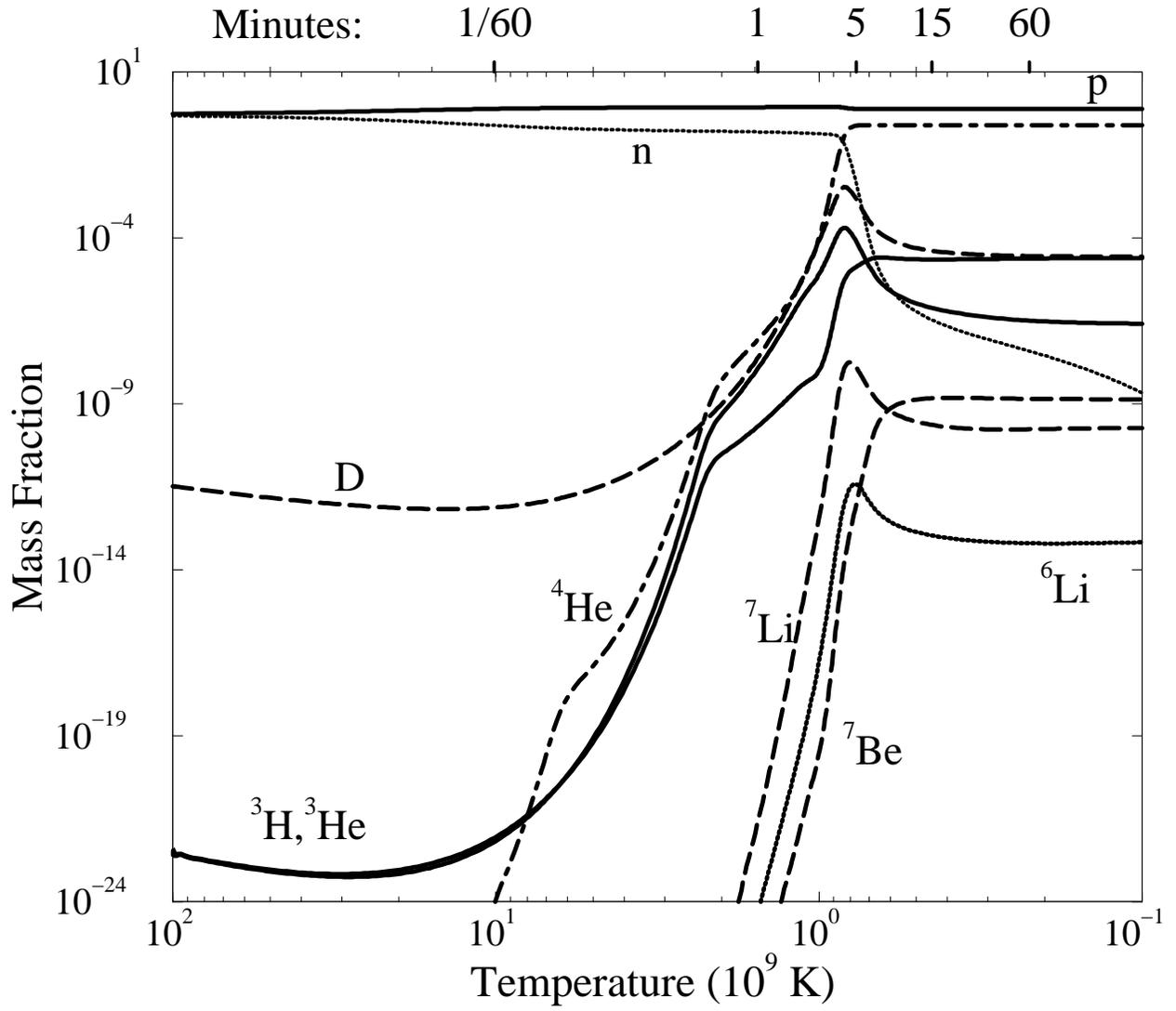,width=150mm,angle=270}
\vspace*{0.5cm}
\caption{Mass fraction of primordial nuclei as a function of
temperature for $\eta = 5.1 \times 10^{-10}$, from 
Burles, Nollett, \& Turner (1999), 
with kind permission from 
Kenneth Nollett.}
\end{figure}

\newpage

Our knowledge of the observed primordial abundances is still
uncertain, allowing and involving different models for
primordial nucleosynthesis. Though the standard cosmology
tells us that all nuclei heavier than carbon were produced in stellar
interiors after the galaxy formation,
recent observations in the
sub-giant CD-38$^{\circ}$245, one of the oldest stars in first
generation which were born a few million years after the
galaxy formation, as well as recently observed absorption lines
in quasi stellar objects (QSO) with red-shift factor $z = 2$
(see, e.g., Kajino 1992) 
%\cite{kajino93}
suggest that there were some 
production activities of 
medium and heavy elements (up to Ba) before or during the
galaxy formation. 
At present,
there are no definite measurements of the primordial 
abundances for carbon and heavier elements, therefore 
all such observations should be
interpreted only as an upper, possible limit.

At the end of this section,
let us note one more point 
of interest in context with the aim of 
the present paper. 
Even if the standard BBN explains the origin of
light elements D, $^3$He, $^4$He, and $^7$Li and their primordial
abundances, it is hoplessly ineffective in generating
$^6$Li, $^9$Be, $^{10}$B, $^{11}$B (see, e.g., Vangioni-Flam,
Casse, \& Audoze 1999). Due to their low binding energy, these
nuclei are not produced significantly in the BBN or in stellar 
nuclear burning,
and are, in fact, destroyed in stellar interiors. 
Instead, it is believed today that LiBeB are made
mostly by spallation processes 
due to energetic nuclei and neutrinos
(see Fields, Olive, Vangiony-Flam, \& Olive 1999 and references therein).
We will return to this question again in Section 5.
Let us also mention that a new,  good, and useful review on
BBN nucleosynthesis and primordial abundances will be published shortly 
in {\em Physica Scripta} by Tytler, O'Meara, Suzuki, \& Lubin (2000).

\section{Nucleosynthesis in Stars}

After the Big Bang, the story of nucleogenesis is considered mostly with
the physics of stellar evolution and nucleosynthesis in stars
(see, e.g., McWilliam 1997). In the B$^2$FH paper, the
``bible" of nuclear astrophysics, 
to describe all features of the abundance curve 
known as of 1957, eight separate processes were necessary to be taken
into account: 
1) Hydrogen Burning; 
2) Helium Burning; 
3) $\alpha$ Process;
4) $e$ Process; 
5) $r$ Process; 
6) $p$ Process; 
7) $s$ Process; and
8) $x$ Process. 
Today, 43 years later, nearly the same 
processes are still considered to be as
fundamental ones for stellar nucleosynthesis (Wallerstein et al. 1997).
For completeness sake, let us briefly list bellow 
in turn processes shown in Fig. 1 which are believed
today to be of the main importance for stellar nucleosynthesis.

\subsection{Hydrogen Burning}
Hydrogen burning starts in stars with the proton-proton and deuteron-proton
reactions (1) and (6) discussed in Section 3. Other reactions of the
hydrogen burning chain 
suggested and discussed half a century ago by many prominent physicists 
(see detailed references in Lang 1999) are (13) and (15), as well as:
\begin{equation}
^3\rm{He} + ^3\rm{He} \to ^4\rm{He} + \rm{p} + \rm{p} , 
\end{equation}
\begin{equation}
^7\rm{Be} + e^- \to ^7\rm{Li} + \nu_e , 
\end{equation}
\begin{equation}
^7\rm{Be} + \rm{p} \to ^8\rm{B} + \gamma , 
\end{equation}
\begin{equation}
^8\rm{B} \to ^8\rm{Be} + e^+ +  \nu_e , 
\end{equation}
\begin{equation}
^8\rm{Be} \to ^4\rm{He} + ^4\rm{He}  . 
\end{equation}
The energy released in each of these and other
reactions discussed 
%bellow 
may be found in
Lang 1999. For stars more massive than the Sun, hydrogen will be 
fused into helium by the fast C-N cycle provided  that carbon, nitrogen,
or oxygen are present to act as a catalyst:
\begin{equation}
^{12}\rm{C} + \rm{p} \to ^{13}\rm{N} + \gamma , 
\end{equation}
\begin{equation}
^{13}\rm{N} \to ^{13}\rm{C} + e^+ + \nu_e , 
\end{equation}
\begin{equation}
^{13}\rm{C} + \rm{p} \to ^{14}\rm{N} + \gamma , 
\end{equation}
\begin{equation}
^{14}\rm{N} + \rm{p} \to ^{15}\rm{O} + \gamma , 
\end{equation}
\begin{equation}
^{15}\rm{O} \to ^{15}\rm{N} + e^+ + \nu_e , 
\end{equation}
\begin{equation}
^{15}\rm{N} + \rm{p} \to ^{12}\rm{C} + ^{4}\rm{He} . 
\end{equation}
Additional proton capture reactions, 
which may take place to form the complete C-N-O bi-cycle
(see references in Lang 1999) are:
\begin{equation}
^{15}\rm{N} + \rm{p} \to ^{16}\rm{O} + \gamma , 
\end{equation}
\begin{equation}
^{16}\rm{O} + \rm{p} \to ^{17}\rm{F} + \gamma , 
\end{equation}
\begin{equation}
^{17}\rm{F} \to ^{17}\rm{O} + e^+ + \nu_e , 
\end{equation}
\begin{equation}
^{17}\rm{O} + \rm{p} \to ^{14}\rm{N} + ^{4}\rm{He} . 
\end{equation}
It is possible that the CNO cycle produces most of the $^{14}$N found in 
nature. During supernovae explosions, a rapid CNO cycle might take place in
which the (n,p) reactions replace the beta decays in the cycle.

\subsection{Helium  Burning}

It is believed now that helium burning results in the production 
of approximately equal amount of $^{12}$C and $^{16}$O in stars of masses 
from 0.5 to 50 $M_\odot$. The reactions assigned to be the triple alpha 
process, $^{4}\rm{He} + ^{4}\rm{He} + ^{4}\rm{He}  \to ^{12}\rm{C} + \gamma$, are 
(see references in Lang 1999):
\begin{equation}
^{4}\rm{He} + ^4\rm{He} \to ^{8}\rm{Be} , 
\end{equation}
\begin{equation}
^{8}\rm{Be} + ^4\rm{He} \to ^{12}\rm{C}^* , 
\end{equation}
\begin{equation}
^{12}\rm{C}^* \to ^{12}\rm{C} + \gamma . 
\end{equation}
Once $^{12}$C is formed, 
and with increasing temperature in the stellar core,
$^{16}$O and other heavier nuclei
up to the very stable, double magic, $^{40}$Ca, or even a little
further
will be produced 
by successively $\alpha$-capture:
\begin{equation}
^{12}\rm{C} + ^4\rm{He} \to ^{16}\rm{O} + \gamma , 
\end{equation}
\begin{equation}
^{16}\rm{O} + ^4\rm{He} \to ^{20}\rm{Ne} + \gamma , 
\end{equation}
\begin{equation}
^{20}\rm{Ne} + ^4\rm{He} \to ^{24}\rm{Mg} + \gamma , 
\end{equation}
\begin{equation}
^{24}\rm{Mg} + ^4\rm{He} \to ^{28}\rm{Si} + \gamma , 
\end{equation}
\begin{equation}
^{28}\rm{Si} + ^4\rm{He} \to ^{32}\rm{S} + \gamma , 
\end{equation}
\begin{equation}
^{32}\rm{S} + ^4\rm{He} \to ^{36}\rm{Ar} + \gamma , 
\end{equation}
\begin{equation}
^{36}\rm{Ar} + ^4\rm{He} \to ^{40}\rm{Ca} + \gamma . 
\end{equation}
As suggested by Cameron about half a century ago (see references in
Cameron 1999 and Lang 1999), $\alpha$-capture reactions on products of the 
C-N-O cycle might play a role of neutron producer in stars:
\begin{equation}
^{13}\rm{C} + ^4\rm{He} \to ^{16}\rm{O} + \rm{n} , 
\end{equation}
\begin{equation}
^{14}\rm{N} + ^4\rm{He} \to ^{18}\rm{F} + \gamma , 
\end{equation}
\begin{equation}
^{18}\rm{F} \to ^{18}\rm{O} + e^+ + \nu_e ,
\end{equation}
\begin{equation}
^{18}\rm{O} + ^4\rm{He} \to ^{22}\rm{Ne} + \gamma , 
\end{equation}
\begin{equation}
^{18}\rm{O} + ^4\rm{He} \to ^{21}\rm{Ne} + \rm{n} , 
\end{equation}
\begin{equation}
^{22}\rm{Ne} + ^4\rm{He} \to ^{25}\rm{Mg} + \rm{n} . 
\end{equation}

\subsection{Carbon and Oxygen Burning}

At the condition of helium burning, the predominant nuclei are $^{12}$C
and $^{16}$O (Lang 1999). When temperatures greater than 
$8 \times 10^8$ 
%$^\circ$
K 
are reached, carbon will begin to react with itself according to
the reactions: 
\begin{equation}
^{12}\rm{C} + ^{12}\rm{C} \to ^{24}\rm{Mg} + \gamma ,
\end{equation}
\begin{equation}
^{12}\rm{C} + ^{12}\rm{C} \to ^{23}\rm{Na} + \rm{p} ,
\end{equation}
\begin{equation}
^{12}\rm{C} + ^{12}\rm{C} \to ^{20}\rm{Ne} + ^4\rm{He} ,
\end{equation}
\begin{equation}
^{12}\rm{C} + ^{12}\rm{C} \to ^{23}\rm{Mg} + \rm{n} ,
\end{equation}
\begin{equation}
^{12}\rm{C} + ^{12}\rm{C} \to ^{16}\rm{O} + ^4\rm{He} + ^4\rm{He} .
\end{equation}
At about 
$2 \times 10^9$ 
%$^\circ$
K, 
oxygen will also react with itself according to the reactions:
\begin{equation}
^{16}\rm{O} + ^{16}\rm{O} \to ^{32}\rm{S} + \gamma ,
\end{equation}
\begin{equation}
^{16}\rm{O} + ^{16}\rm{O} \to ^{31}\rm{P} + \rm{p} ,
\end{equation}
\begin{equation}
^{16}\rm{O} + ^{16}\rm{O} \to ^{31}\rm{S} + \rm{n} ,
\end{equation}
\begin{equation}
^{16}\rm{O} + ^{16}\rm{O} \to ^{28}\rm{Si} + ^4\rm{He} ,
\end{equation}
\begin{equation}
^{16}\rm{O} + ^{16}\rm{O} \to ^{24}\rm{Mg} + ^4\rm{He} + ^4\rm{He} .
\end{equation}

The $\alpha$ particles, protons, and neutrons which are produced via
reactions (52-61) will interact with the other products of the burning to
form many other nuclides with $16 \le A \le 28$. 

It is now thought that most of the carbon, oxygen, and silicon burning,
which account for the observed solar system abundances for $20 \le A \le 64$,
occurs during fast explosions, and these explosive burning processes
are discussed briefly in Section 4.7.

\subsection{Silicon Burning}

At the completion of carbon and oxygen burning, the most abundant nuclei 
will be $^{32}$S and $^{28}$Si with significant amount of $^{24}$Mg
(Lang 1999).
Because the binding energies for protons, neutrons, and $\alpha$ particles
in $^{32}$S are smaller than those in $^{28}$Si, the nuclide
$^{32}$S will be the first to photodisintegrate according to the reactions:
\begin{equation}
^{32}\rm{S} + \gamma \to ^{31}\rm{P}  + \rm{p} ,
\end{equation}
\begin{equation}
^{31}\rm{P} + \gamma \to ^{30}\rm{Si} + \rm{p} ,
\end{equation}
\begin{equation}
^{30}\rm{Si} + \gamma \to ^{29}\rm{Si} + \rm{n} ,
\end{equation}
\begin{equation}
^{29}\rm{Si} + \gamma \to ^{28}\rm{Si} + \rm{n} .
\end{equation}
The resulting reactions will leave little but $^{28}$Si. Silicon will then 
begin to photodisintegrate at temperatures greater than
$3 \times 10^9$ $^\circ$K 
according to the reactions:
\begin{equation}
^{28}\rm{Si} + \gamma \to ^{27}\rm{Al} + \rm{p} ,
\end{equation}
\begin{equation}
^{28}\rm{Si} + \gamma \to ^{24}\rm{Mg} + ^4\rm{He} .
\end{equation}
As the $(\gamma,^4\rm{He})$ reaction has the lower threshold, it is the
dominant reaction at low temperatures
$ T < 2 \times 10^9$ $^\circ$K;
whereas the $(\gamma,\rm{p})$ reaction has the shorter lifetime at higher 
temperatures. Further photodisintegrations lead to the build-up
of lighter elements according to the reactions: 
\begin{equation}
^{24}\rm{Mg} + \gamma \to ^{23}\rm{Na} + \rm{p} ,
\end{equation}
\begin{equation}
^{24}\rm{Mg} + \gamma \to ^{20}\rm{Ne} + ^4\rm{He} ,
\end{equation}
\begin{equation}
^{20}\rm{Ne} + \gamma \to ^{16}\rm{O}  + ^4\rm{He} ,
\end{equation}
\begin{equation}
^{16}\rm{O}  + \gamma \to ^{12}\rm{C}  + ^4\rm{He} .
\end{equation}

The abundances of most of the nuclei in the range $28 \le A \le 60$
are thought to be determined by equilibrium or quasi-equilibrium processes
in which the importance of many individual reaction rates is diminished
(see references in Lang 1999).
Most nuclear species between $^{28}$Si and $^{59}$Co,
except the neutron-rich species ($^{36}$S,  $^{40}$Ar,  
$^{43}$Ca,  $^{46}$Ca,  $^{48}$Ca,  $^{51}$Ti,  $^{54}$Cr, and $^{58}$Fe),
are generated by a quasi-equilibrium process in which the only
important thermonuclear reaction rates are thought to be those
of $^{44}$Ca, $^{45}$Sc, and $^{45}$Ti (Lang 1999).
The abundances of the neutron-rich species could be determined by the
$s$ or $r$ processes discussed briefly bellow.

\subsection{{\em s}, {\em r}, and {\em p} Processes}

Because the binding energy per nucleon decreases with increasing $A$
for nuclides beyond the iron peak $(A \ge 60)$, and because these elements 
have large Coulomb barriers, they are not likely to be formed by fusion or 
alpha and proton capture
(Lang 1999).
It is thought that most of these elements are formed by neutron
capture reactions which start with the iron group nuclei
(Cr, Mn, Fe, and Ni). If the flux of neutrons is week, most chains of 
neutron capture will include only a few capture before the beta decay
of the product nucleus. As the neutron capture lifetime is slower 
{\bf ({\em s})}
than the beta decay lifetime, this type of neutron capture is called the
{\bf {\em s} process}. This process can continue all the way up to lead and 
bismuth; beyond bismuth the resulting nuclei alpha decay back to Pb and
Tl isotopes (Wallerstein et al. 1997). Good reviews on laboratory 
measurements, stellar models, and abundance studies of the $s$-process
elements may be found in Secs. X and XI of the recent comprehensive
surveys by Wallerstein et al. (1997) and in  K\"appeler (1999).

When there is a strong neutron flux, as it is believed to occur during 
a supernovae explosion, the neutron-rich elements will be formed by 
the rapid {\bf ({\em r})} neutron capture process, in which the sequental 
neutron captures take place on a time scale which is much more shorter
than for beta decay of the resulting nuclei. 
This process produces  much more neutron-reach progenitors
that are required to account for the second set of abundance peaks that 
are observed about 10 mass units above the $s$-process abundance
peaks corresponding to the neutron magic numbers, $N = 50$ and 82.
We forward readers interested in more details about
both the physics and astrophysical 
scenario of the rapid neutron capture to the Sec. XII of the mentioned
above review by 
Wallerstein et al. (1997) and to a more recent and useful work by
Cowan et al. (1999).

The proton rich medium and
heavy elements are much less abundant than the elements
thought to be produced by $r$ and $s$ processes, and are thought to be 
formed by a proton capture {\bf ({\em p} process)} at high enough temperature to
overcome the coulomb barrier. 
Burbidge, Burbidge, Fowler, \& Hoyle (1957)
described in their ``bible",
two possible mechanisms by
which $p$-nuclides could be formed: proton radiative captures, 
$(\rm{p},\gamma)$,
in a hot 
($ T \sim  2-3 \times 10^9$ $^\circ$K)
proton-rich environment, or photon-induced
n, p, and $\alpha$-particle removal reactions,
also in a hot environment. 
A possible occasion for this process is the passage of a supernova shock wave 
through the hydrogen outer layer of a pre-supernova star.
The separate mechanisms that are believed today as
contributing to $p$-process nucleosynthesis, as 
well as their strengths and weaknesses are discussed
in details in Sec. XIV of the review by Wallerstein et al. (1997).

It is believed today that some nucleosynthesis of the lighter
$p$-nuclides is provided by the so called {\bf {\em rp} process}. 
The $rp$-process
is very similar to the $r$-process, except it goes by
a successive rapid proton absorption and $\beta^+$ decay. At
present, it is believed that the $rp$-process can provide contributions
to the nucleosynthesis of proton rich isotopes after the hot C-N-O cycle
up through $^{65}$As, to as high as $^{68}$Se, or even to $^{96}$Ru
(see details and references in Wallerstein et al. 1997). 

\subsection{Equilibrium Processes}

Another type of processes 
of nucleosynthesis in stars
discussed intensively in the literature
since the pioneering work by Hoyle (1946) and reviewed in B$^2$FH are the
{\bf equilibrium processes}, called in B$^2$FH as ``$e$ processes".
Such processes are possible only if the matter is in equilibrium with the 
radiation, and if every nucleus is transformable into any other nucleus. 
Hoyle (1946) showed that matter is in equilibrium with radiation 
at temperatures $T \approx 10^9$ $^\circ$K, and that all known nuclei may 
be transformed into any other nucleus by nuclear reactions at
$T \gtrsim 2 \times 10^9$ $^\circ$K. Though statistical equilibrium requires
that the entropy of a system should be at the maximum, that may be
a too strong requirement, not
fulfilled exactly for real systems (Lang 1999; Wallerstein et al. 1997),
this method proved to be very successful for the description of abundances
of nuclei in the iron group and around $(28 \le A \le 60)$ (see detailed
references in Lang 1999). What is more, if to assume a thermodynamic
equilibration in a star, than its composition (elemental abundances) may
be calculated without determining
individual reaction rates, and only the binding energies 
and partition functions of the various
nuclear species need to be specified.

Under conditions of statistical equilibrium, the number density, $N_i$, of
particles of the $i$th kind is given by (Lang 1999):
\begin{equation}
N_i = {1\over V} 
\sum^n_r \mu_i [\pm \mu_i + \exp (\varepsilon_{ir} /kT)]^{-1} ,
\end{equation}
where $V$ is the volume, $\mu_i$ is the chemical potential of the $i$th particle, 
the plus and minus signs refer to Fermi-Dirac and Bose-Einstein statistics, 
respectively, and the summation is over all energies, $\varepsilon_{ir}$,
which includes both internal energy levels and the kinetic energy. If an
internal level has spin, $J$, then $2J + 1$ states
of the same energy must be included
in the sum. When the nuclides are non-degenerate and non-relativistic, 
Maxwellian statistics can be employed to give (Lang 1999):
\begin{eqnarray}
N_i &=& {\mu_i\over V} 
\Biggl[
\sum (2J_r + 1) \exp \Biggl(- {\varepsilon_{r}\over kT}\Biggr)
\Biggr] \Biggl[{4 \pi V\over h^3} \int\limits^p_0 p^2 
\exp \Biggl( - {p^2\over 2M_ikT} \Biggr) dp \Biggr] = 
\nonumber\\
&=& \mu_i \omega_i \Biggl( {2\pi M_i k T\over h^2} \Biggr) ^{3/2} ,
\end{eqnarray}
where $p$ is the particle momentum, $M_i$ is its mass,
the partition function $\omega_i = \sum (2J_r+1)\exp(-\varepsilon_r /kT)$, 
and here $\varepsilon_r$ refers to internal states only.
For particles $p_i, p_j, \cdots$ which react according to
\begin{equation}
\alpha p_i + \beta p_j + \cdots \leftrightarrows \xi p_r + \eta p_s + \cdots ,
\end{equation}
the chemical potentials are related by the equation
\begin{equation}
\mu^\alpha_i \mu^\beta_j  \cdots = \mu^\xi_r \mu^\eta_j  \cdots \exp(-Q/kT) ,
\end{equation}
where
\begin{equation}
Q = c^2 [\alpha M_i + \beta M_j + \cdots - \xi M_r -\eta M_j - \cdots ] .
\end{equation}

Hoyle (1946) and Burbidge, Burbidge, Fowler, \& Hoyle (1957) considered the
condition of statistical equilibrium between the nuclei, $(A,Z)$,
and free protons, p, and neutrons, n. For a nucleus, there are
$Z$ protons and $(A-Z)$ neutrons and the statistical weight of both protons 
and neutrons is two. It then follows from Eqs. (73) to (76) that for 
equilibrium between nuclides, protons, and neutrons, the number density, 
$N(A,Z)$, of the nucleus, $(A,Z)$, is given by:
\begin{equation}
N(A,Z) = \omega(A,Z)  \Biggl({AM_\mu kT\over 2\pi \hbar^2} \Biggr)^{3/2}
\Biggl( {2\pi \hbar^2\over M_\mu k T}\Biggr)^{3A/2}
{N_{\rm{n}}^{(A-Z)}N_{\rm{p}}^Z\over 2^A} \exp \Biggl[{Q(A,Z)\over kT} \Biggr] , 
\end{equation}
where the partition function, $\omega(A,Z)$, of the nucleus,
$(A,Z)$, is given by
\begin{equation}
\omega(A,Z) = \sum_r (2I_r+1) \exp \Biggl(- {E_r\over k T}\Biggr) ,
\end{equation}
where $I_r$ and $E_r$ are, respectively, the spin and energy of the
$r$th excited level, the binding energy, $Q(A,Z)$, of
the nucleus, $(A,Z)$, is given by
\begin{equation}
Q(A,Z) =c^2[(A-Z)M_{\rm{n}} + ZM_{\rm{p}} - M(A,Z)] ,
\end{equation}
where $M_{\rm{n}}$, $M_{\rm{p}}$, and $M(A,Z)$ are, respectively, the masses
of the free neutron, free proton, and the nucleus, $(A,Z)$, the factor
\begin{equation}
\Biggl({2\pi\hbar^2\over M_\mu k T}\Biggr)^{3/2} \approx 1.6827 \times 
10^{-34} T_9^{-3/2} \mbox{ cm}^{-3} ,
\end{equation}
where $T_9 = T/10^9$,
the atomic mass unit is $M_\mu$, and $N_{\rm{n}}$ and $N_{\rm{p}}$ denote, respectively,
the number densities of free neutrons and protons.
As one can see, Eq. (77) contains indeed only the binding energy
$Q(A,Z)$ and does not require any cross sections or nuclear rates.
Further details, more references, and newer and more general notions
on equilibrium processes may be found,
e.g., in Lang 1999 and Wallerstein et al. 1997.

There is an allied process to the equilibrium nucleosynthesis,
the so called ``{\bf quasi-equilibrium}" 
process, when the total number of nuclei
in different ranges of atomic number or mass number might be slowly varying
and we may see only a quasi-equilibrium between nuclides of some
separate groups, but not between different groups.
So, Michaud and Fowler (1972) showed that 
with an initial neutron enhancement of $4 \times 10^{-3}$
the natural abundances for nuclei with $28 \lesssim A \lesssim 59$
may be accounted for by quasi-equilibrium burning. In this case,
a quasi-equilibrium between elements
with $24 \le A \le 44$ and a separate equilibrium for elements with
$46 \le A \le 60$ is assumed, and detailed nuclear reactions 
are given for the ``bottleneck" at $A=45$ 
(see further details and references on quasi-equilibrium
processes in
Lang 1999 and Wallerstein et al. 1997).
This quasi-equilibrium
silicon burning process must have taken place in a 
short time, $t \lesssim 1$ sec, and at high temperatures,
$T \ga 4.5 \times 10^{9}$ $^\circ$K, suggesting the explosive 
burning processes discussed briefly in the next subsection.

\subsection{Explosive Burning Processes}

As explained by Burbidge, Burbidge, Fowler, and Hoyle (1957), the
successive cycles of static nuclear burning and contraction, which
successfully account for much stellar evolution, must end when the
available nuclear fuel is exhausted (Lang 1999). B$^2$FH showed
that the unopposed action of gravity in a helium exhausted stellar core 
leads to violent instabilities and to rapid thermonuclear reactions
in the stellar envelope. Later, Arnett (1968) showed that when cooling by 
neutrino emission in a highly degenerate gas is considered, the
$^{12}$C + $^{12}$C reaction will ignite explosively at core density of 
about $2 \times 10^{9}$ g cm$^{-3}$.
The stellar material is instantaneously heated and then expands adiabatically 
so that the density, $\rho$, and temperature, $T$, are related by
\begin{equation}
\rho(t) \varpropto [T(t)]^3  ,
\end{equation}
for a $\Gamma_3 = 4/3$ adiabat, and a time variable, $t$.
The appropriate time is the hydrodynamic time scale, $\tau_{HD}$, 
given by (Lang 1999) 
\begin{equation}
\tau_{HD} 
%= (24\pi G \rho)^{-1/2} 
\approx 446 \rho^{-1/2} \mbox{ sec.}
\end{equation}
The initial temperature and density must be such that the 
mean lifetime, $\tau_R$, for a nucleus undergoing an
explosive reaction, $R$, must be close to $\tau_{HD}$. For
the interaction of nucleus 1 with a nucleus 2,
\begin{equation}
\tau_{R} = \tau_{2}(1) = 
[N_2 <\sigma v>]^{-1} = \Biggl[\rho N_A {X_2\over A_2} <\sigma v>\Biggr]^{-1} ,
\end{equation}
where the mass density is $\rho$, the $X_2$,
$A_2$, and $N_2$ are, respectively, the mass fraction,
mass number, and number density of nucleus 2, and
$N_A<\sigma v>$ is the reaction rate.
Arnett (1969) used a mean carbon nucleus lifetime,
\begin{equation}
\log \tau_{^{12}C} \approx 37.4 T^{-1/3}_9 - 25.0 - \log_{10} \rho \approx 
\log_{10} \tau_{HD}
\end{equation}
for carbon burning to determine the initial condition of explosive carbon
burning. Knowing the reaction rates, Eqs. (81) and (84) allow
us to calculate expected abundances using the 
corresponding abundance equations
discussed briefly bellow. 
Abundance rations which closely
approximate those of the solar system were found for
$^{20}$Ne, $^{23}$Na, $^{24}$Mg, $^{25}$Mg, $^{26}$Mg, $^{27}$Al, $^{29}$Si,
and  $^{30}$Si,
when it was assumed that a previous epoch of helium burning produced 
equal amounts of $^{12}$C and $^{16}$O, and that
\begin{eqnarray}
T_{\rm{p}} &=& 2 \times 10^{9} \mbox{ } ^\circ{\mbox K} \nonumber\\
\rho_{\rm{p}} &=& 1 \times 10^5 \mbox{ g cm}^{-3} \\
\eta &=& 0.002\mbox{ .}\nonumber  
\end{eqnarray}
Here $T_{\rm{p}}$ and $\rho_{\rm{p}}$ denote, respectively,
the peak values of temperature and mass density in the shell under 
consideration, and the neutron excess, $\eta$, is given by
\begin{equation}
\eta = {N_{\rm{n}} - N_{\rm{p}}\over N_{\rm{n}} + N_{\rm{p}}} ,
\end{equation}
where $N_{\rm{n}}$ and $N_{\rm{p}}$ 
denote, respectively, the number density of free 
neutrons and protons (Lang 1999).

Similarly, many works by different authors were dedicated to study
explosive oxygen and silicon burning. Useful references and more details
on explosive nucleosynthesis may by found in the comprehensive
monograph by Lang (1999) and in the recent reviews by Arnett (1995) and
Woosley and Weaver (1995).

In a general case, the equation governing the change in the number
density, $N(A,Z)$, of the nucleus $(A,Z)$ is of the form (Lang 1999):
\begin{equation}
{d\over dt}(N_i) = - \sum_j N_i N_j <\sigma v>_{ij}
+ \sum_{kl} N_k N_l <\sigma v>_{kl} \mbox{ ,}
\end{equation}
where $N_i$ is the number density of the $i$th species, $<\sigma v>_{ij}$
is the product of cross section and the relative velocity for an interaction
involving species $i$ and $j$, the
$N_m N_n$ is replaced by $N_n^2 /2$ for identical particles, and the 
summation
is over all reactions which either create or destroy the species, $i$. 
The probabilistic interpretation of this equation is obvious: the number
density, $N_i$ of the species $i$ at a given time $t$ is built by all
processes resulting in the species of interest minus the contribution
of all processes destroying these species. In practice, for numerical
calculations, instead of $N_i$
one usually uses the following parameter (Lang 1999):
\begin{equation}
Y_i = Y(A,Z) = {N(A,Z)\over \rho N_A} = {N_i\over \rho N_A} ,
\end{equation}
where $\rho$ is the mass density of the gas under consideration and
$N_A$ is the Avogadro's number. Then, Eg. (87) can be rewritten as:
\begin{equation}
{d\over dt} (Y_i) = - \sum_j f_{ij} + \sum_{kl} f_{kl} ,
\end{equation}
where the vector flow, $f_{ij}$, which contains nuclei $i$ and $j$ in
the entrance channel, is given by
\begin{equation}
f_{ij} = {N_i N_j <\sigma v>_{ij}\over \rho N_A} = 
Y_i Y_j \rho N_A <\sigma v>_{ij} ,
\end{equation}
The real, explicit view of this equation in a concrete calculation
depends on processes we like to 
take into account. So, in a general case when we take into account
negative and positive $\beta$-decays, electron and neutrom captures, 
alpha decay, photodisintegration, as well as all possible reactions between
two interacting nuclei, Eq. (89) becomes (Lang 1999):
\begin{eqnarray}
{dY(A,Z)\over dt}  = &-& [\lambda_{\beta^-}(A,Z) + 
\lambda_{\beta^+}(A,Z) + \lambda_{K}(A,Z) + \lambda_{\alpha}(A,Z)+ \nonumber\\
&+& \lambda_{\gamma}(A,Z) + 2.48 \times 10^8 \sigma _T N_{\rm{n}} +
\sum_j Y(A_j,Z_j) \rho N_A <\sigma v>_j ] Y(A,Z) + \nonumber\\
&+& \lambda_{\beta^-}(A,Z-1)Y(A,Z-1) +  \lambda_{\beta^+}(A,Z+1)Y(A,Z+1) + \\
&+& \lambda_{K}(A,Z+1)Y(A,Z+1) +  \lambda_{\alpha}(A+4,Z+2)Y(A+4,Z+2) +
\nonumber\\
&+& \lambda_{\gamma}(A,Z)Y(A,Z) + 2.48 \times 10^8 \sigma _T N_{\rm{n}} 
Y(A-1,Z) +
\nonumber\\
&+&\sum_{ik} Y(A_i,Z_i) Y(A_k,Z_k) \rho N_A <\sigma v>_{ik}\nonumber\mbox{ ,}\\
\end{eqnarray}
where the symbol $\lambda$ denotes the decay rate or the inverse mean 
lifetime, the subscritps $\beta^-$, $\beta^+$, $K$, $\alpha$,
and $\gamma$ denote, respectively, negative beta decay, positive
beta decay, electron capture, alpha decay, and photodisintegration, 
$\sigma_T$ is the cross section for neutron capture in cm$^2$,
$N_{\rm{n}}$ is the number density of neutrons, the summation $\sum\limits_j$ 
denotes all 
reactions between the nucleus (A,Z) and any other nucleus, 
the summation $\sum\limits_{ik}$ denotes all reactions
between two nuclei which have $(A,Z)$ as a product, $\rho$ is the gas 
mass density, and $N_A<\sigma v>$ is the reaction rate.

Numerical solution of such complex set of nuclear
reaction networks requires a number of approximations and assumptions.
Details on abundance equations for $s$, $r$, equilibrium, and
quasi-equilibrium processes as well as useful references can be
found in Lang (1999).
 
%\section{$\nu$-Processes}

\section{Li-Be-B Generation and Spallation Processes}

%\subsection{LiBeB generation}

The observed today rare light nuclei, lithium, beryllium, 
and boron, are not products of the BBN or stellar nucleosynthesis,
and are, in fact destroyed in hot stellar interiors. 
This condition is reflected in the comparatively low abundances of these 
nuclei (see Figs. 1 and 2).
%
%As mentioned in Sec. 3, 
%due to the fact that nuclei with mass 5 and 8 are unstable, 
%the BBN has practically stopped 
%at A$ = 7$ and primordial nucleosynthesis has been unable to proceed 
%efficiently beyond lithium 
%\cite{VangioniFlam99}.
%(Vangioni-Flam, Cass\'e, \& Audouze 1999).
%The Big Bang production of $^6$Li is dominated by the 
%D$(\alpha, \gamma)^6$Li reaction 
%\cite{smith93,VangioniFlam99}.
%(Smith, Kawano, , \& Malaney 1993;
%Vangioni-Flam, Cass\'e, \& Audouze 1999).
%It is believed that the BBN $^6$Li production was tiny
%(in addition, stellar thermonuclear processes destroy $^6$Li),
%its abundances is unobservably small, thus the most of the observed
%$^6$Li could be produced in Galactic, non-equilibrium processes
%of cosmic ray interactions
%(e.g. Fields \& Olive 1998).
In contradiction with measurements,
the primordial
$^6$Li,
$^9$Be, $^{10}$B, and $^{11}$B abundances 
calculated 
using the best
of the available today evaluations for the reaction rates are many orders
of magnitude below compared to $^7$Li,
so, the standard BBN is 
%hoplessly 
ineffective in generating
$^6$Li,
$^9$Be, $^{10}$B, and $^{11}$B 
%\cite{VangioniFlam99}.
(Vangioni-Flam, Cass\'e, \& Audouze 1999).
 
Up to recently, the most plausible formation agents of LiBeB were
thought to be Galactic Cosmic Rays (GCRs) interactions with 
interstellar medium (ISM), mainly
C, N, and O nuclei. (The most abundant and energetic cosmic-ray
particles are protons and $\alpha$-particles.)
Other possible origins have been also identified:
primordial and stellar ($^7$Li) and supernova neutrino spallation
(for $^7$Li and $^{11}$B), while 
$^6$Li, $^9$Be, and $^{10}$B are thought to be pure spallation
products 
%\cite{VangioniFlam99}.
(Vangioni-Flam, Cass\'e, \& Audouze 1999).

Recent measurements in a few halo stars
with the 10 meter KECK telescope and the Hubble Space Telescope 
indicate a quasi linear correlation between Be and B vs Fe,
at least at low metallicity,
contradictory at first sign to a dominating GCRs origin of the light elements
which predicts a quadratic relationship
(see the appendix in 
Vangioni-Flam, Cass\'e, \& Audouze 1999).
%\cite{VangioniFlam99}).
As a consequence, the theory of the origin and evolution
of the LiBeB nuclei has yet to be reassessed
Vangioni-Flam, Cass\'e, \& Audouze 1999).
Aside GCRs, which are thought to be accelerated in the general 
interstellar medium and which create Li-Be-B through the break up 
of interstellar C-N-O nuclei by their fast
protons and alphas, Wolf-Rayet stars (WR) and core collapse supernovae (SNII)
grouped on superbubbles could produce copious amount of light elements via 
the fragmentation in flight of rapid carbon and oxygen nuclei
(called hereafter low energy component, LEC) colliding with H and He
in the ISM
(Vangioni-Flam, Cass\'e, \& Audouze 1999).
In this case, Li-Be-B would be produced independently of the interstellar 
medium chemical composition. As noted by 
Vangioni-Flam, Cass\'e, \& Audouze (1999),
more spectroscopic observations (specifically of O, Fe, Li, Be, B)
in halo stars are required for a better understanding of the relative 
contribution of various mechanisms.

New measurements of Be/H and B/H, together with [Fe/H]
(see detailed references in
Vangioni-Flam, Cass\'e, \& Audouze 1999)
in very low metallicity halo stars came to set strong constrains on the
origin of light isotopes. 
Recent 
compilations of 
Be and B data are presented in Figs. 7 and 8. The most striking point is 
that log(Be/H) and log(B/H) are both quasi proportional
to [Fe/H]. 

\newpage
\begin{figure}[h!]
%\vspace*{-5.0cm}
\hspace*{5mm}
\psfig{figure=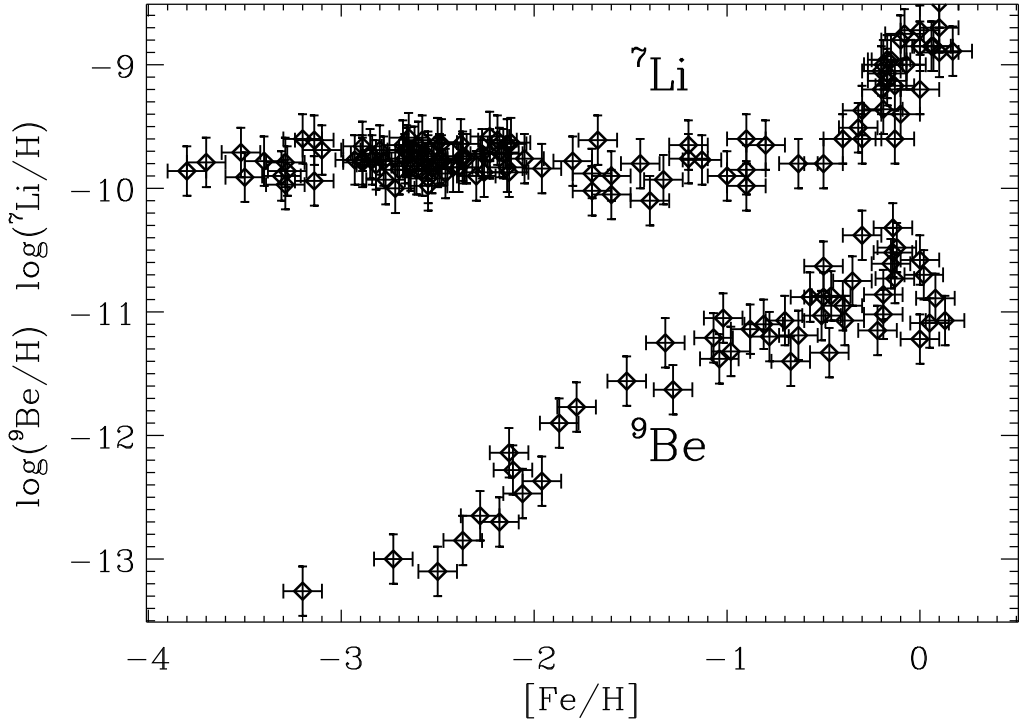,width=150mm,angle=0}
%\vspace*{0.5cm}

\caption{
Lithium (essentially $^{7}$Li) and Beryllium abundances as a function 
of the iron abundance of stars
(Vangioni-Flam, Cass\'e, \& Audouze 1999).
This figure is reproduced from 
%reference
Vangioni-Flam, Cass\'e, \& Audouze (1999) with kind permission
by Elisabeth Vangioni-Flam.
}
\end{figure}
{\noindent This linearity came as a surprise since a quadratic relation 
was expected from the GDR mechanism, whereas the suprnovae
origin would lead naturally to slope 1. }
This was a strong indication 
that the standard GCRs are not the main producers of Li-Be-B in the
early Galaxy
(Vangioni-Flam, Cass\'e, \& Audouze 1999). 

Concerning lithium, as one can see from Fig. 7, the 
flat portion of the lithium abundance, usually referred
to as the Spite plateau 
(after the original work of Francois and Monique Spite in 1982)
expends 
up to [Fe/H] $\sim$ -1. It is believed that it represents
the abundance of Li generated by the BBN nucleosynthesis.
Beyond, Li/H is strongly
increasing until its solar value of $2 \times 10 ^{-9}$. 
This increase in the Li/H ratio is believed to be related with 
nucleosynthesis in a variety of Galactic objects, including
Type II supernovae, novae and giant stars, as well as 
production by cosmic rays (Ramaty, Kozlovsky, \& Lingenfelter 1998).
A stringent constraint on any theory of Li evolution 
arises from such a form of Li/H curve:
it should avoid to cross 
the Spite's plateau below [Fe/H] = -1. Accordingly, the Li/Be production
ratio should be less than about 100
(Vangioni-Flam, Cass\'e, \& Audouze 1999). 
 
Galactic cosmic rays represent the only sample of matter originating from
beyond the Solar System. They are constituted by bare nuclei stripped from 
their electrons. Their energy density 
%{\noindent 
(about 1 eV cm$^{-3}$ similar
to that of stellar light and that of galactic magnetic field), 
indicate that they are an important component in the dynamics of the
Galaxy
(Vangioni-Flam, Cass\'e, \& Audouze 1999). 
%}
A key point for us is that, as can be seen from Fig. 2,
GCRs are exceptionally LiBeB rich (LiBeB/CNO $\sim$ 0.25)
compared to the Solar System matter (LiBeB/CNO $\sim 10^{-6}$).

%\newpage
\begin{figure}[h!]
%\vspace*{-5.0cm}
\hspace*{5mm}
\psfig{figure=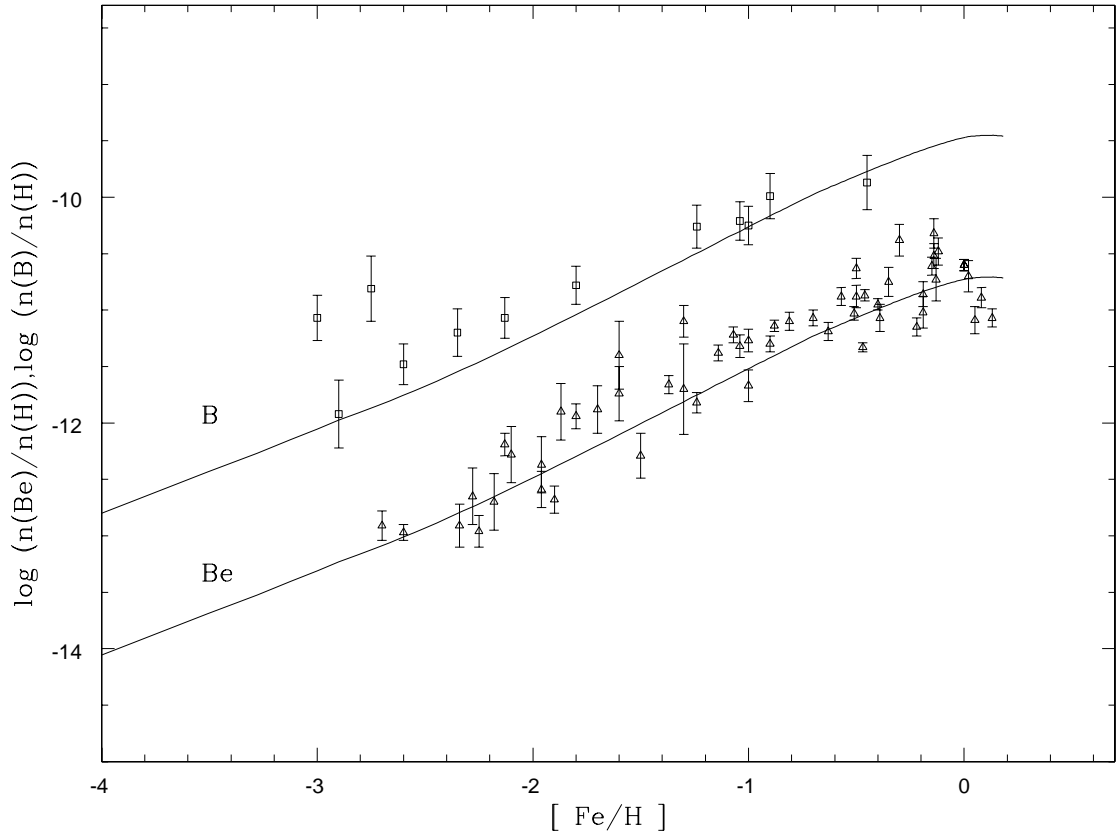,width=120mm,angle=270}
%\vspace*{0.5cm}

\caption{
Beryllium and Boron evolution vs [Fe/H]. The halo evolution,
([Fe/H] $<$ -1), is dominated by the LEC component linked to massive stars. 
As far as B is concerned, there is room for a small contribution for 
$\nu$ spallation.
This figure is reproduced from 
%reference
Vangioni-Flam, Cass\'e, \& Audouze (1999) with kind permission
by Elisabeth Vangioni-Flam.
}
\end{figure}

%\newpage

For detailed calculations
of LiBeB production by the GCR mechanism, 
the formation rate of an light isotope
(i.e., Li, Be, or B, noted here as $L$) from the spallation of an medium
isotope 
(e.g., $^{12}$C, $^{14}$N, $^{16}$O, and $^{20}$Ne, noted as $M$)
by a flux of protons with energy spectrum, $\varphi(E)$, is given
by (Lang 1999):
\begin{equation}
{dN_L\over dt} = \sum_M N_M \int \sigma(M,L,E) \varphi(E) dE ,
\end{equation}
where $M$ denotes any of ``medium" elements and $L$, any of ``light"
nuclei, the number densities of the $M$ and $L$ elements are, respectively, 
$N_M$ and $N_L$, the time variable is $t$, the proton energy is $E$, and 
the spallation reaction cross section is $\sigma(M,L,E)$. For the
low energy cosmic (LEC) rays mechanism, where LiBeB are produced
by interaction of low energy (less than 100 MeV/A) interactions of
``medium" nuclei with interstellar H and $^4$He, we have just
a similar formula. The main difference is in the energy dependence
of fluxes of the projectiles, while the values of cross sections are
the same, for identical bombarding energies per nucleon. For spallative
nucleosynthesis calculations, cross sections at energies from
$\sim 1$ MeV to $\sim 100$ GeV are required, in contrast with the stellar
nucleosynthesis that occur 
at low energies, from $\sim 1$ keV to $\sim 100$ keV.
Current assumptions about energy dependences of the GCR and LEC fluxes,
as well as further
interesting points of the LiBeB story may be found in 
Vangioni-Flam, Cass\'e, \& Audouze (1999) and in proceedings of the 
recent special conference on LiBeB held in December 1998 in Paris
(Ramaty, Vangioni-Flam, Cass\'e, \& Olive 1999).
Spallation cross sections are discussed in the next section.

According to modern concepts (see, e.g.,
%\cite{woosley90,woosley95}), 
Woosley et al. 1990; Woosley \& Weaver 1995),
{\bf neutrino spallation} (NS) is also a source of 
$^7$Li and $^{11}$B via the interaction of neutrinos (predominantly 
$\nu_\mu$ and $\nu_\tau$) on nuclei, specifically on $^4$He and $^{12}$C
%\cite{VangioniFlam99}). 
(Vangioni-Flam, Cass\'e, \& Audouze 1999).
Recently, $\nu$-process nucleosynthesis was 
incorporated 
into a model of galactic chemical evolution
(Olive et al. (1994)
%\cite{olive94}
 and Vangioni-Flam et al. (1996)
%\cite{VangioniFlam96}
which had included as well the LEC component) with the
primary purpose of augmenting the low value for $^{11}$B/$^{10}$B
produced by standard GCR nucleosynthesis.
To fit the observed ration of 4, it was found that the yields
of NS
predicted by 
Woosley \& Weaver (1995)
had to be turned down by a factor of about 2 to 5, to avoid the
overproduction of $^{11}$B. Turning down the NS yields ensured as well 
that the  production of $^7$Li was insignificant, in accordance
with the Spite plateau 
(Vangioni-Flam, Cass\'e, \& Audouze 1999). 

Note that if taking the full NS yield, all galactic boron would be
produced by $\nu$ spallation. This could be a problem since $^9$Be is not
coproduced and $^7$Li overproduced.Thus, the NS mechanism acts as a 
complement to nuclear spallation at a level estimated to at most 20 percent 
concerning $^{11}$B, if one wants to fulfil the observational constraints
of LiBeB discussed by 
Vangioni-Flam et al. (1999).
%\cite{VangioniFlam99}.

An example of contribution from primordial, galactic cosmic rays,
and $\nu$-nucleosynthesis to the total Li abundance, as calculated by
Ryan, Beers, Olive, Fields, \& Norris (1999) is shown in Fig. 9.
Although these results were obtained not without fitting parameters
(therefore are not completely definitive), they may help us to understand
the relative role of different production mechanisms of light elements.
One can see that the primordial contribution to Li abundance 
decreases at high metallicity due to astration, but other components
increase with metallicity as discussed by
Ryan, Beers, Olive, Fields, \& Norris (1999).	
The main conclusion from these results as well as from     
recent works by other
authors (see details and references in
Ryan, Beers, Olive, Fields, \& Norris 1999) is that LiBeB evolution
may be understood only if we take into account a combination
of BBN, cosmic ray, and $\nu$-process nucleosyntheses, but the
$\nu$-process scenario seems to not play a major role.

\begin{figure}[h!]
%\vspace*{+4.0cm}
\hspace*{5mm}
\psfig{figure=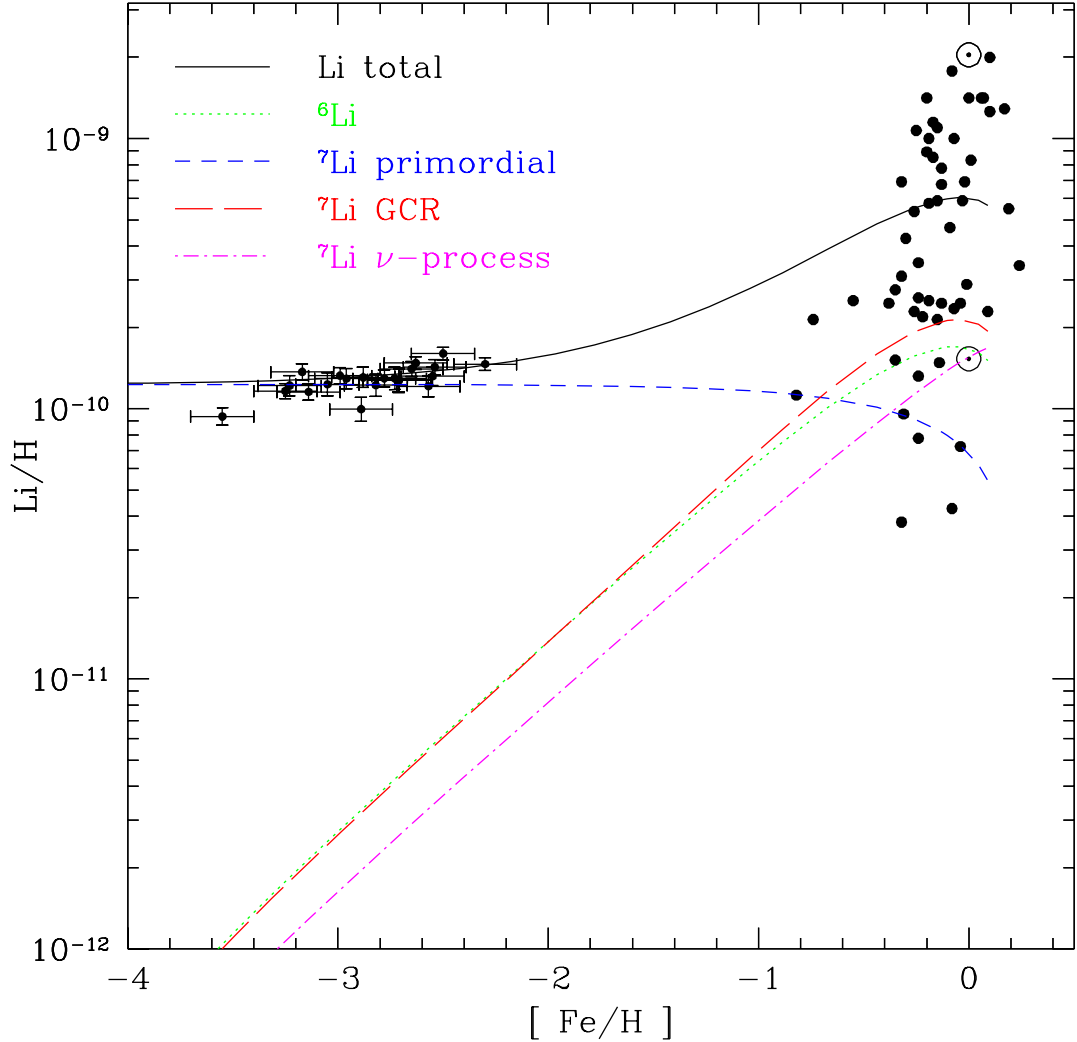,width=150mm,angle=0}
%\vspace*{0.5cm}

\caption{
Contributions to the total 
lithium abundance from different reaction mechanisms shown on the plot, as
predicted 
by the one-zone (closed box) GCE model
(Fields \& Olive 1999)
compared with 
available experimental data for 
low metallicity and
high metallicity stars (Ryan, Beers, Olive, Fields, \& Norris 1999). 
The solid curve is the sum of all 
components;
$^6$Li is thought to be produced only by spallation reactions
(Fields \& Olive 1998). 
This figure is taken with the kind permission of authors
from 
Ryan, Beers, Olive, Fields, \& Norris (1999), where further details
may be found. 
}
\end{figure}
%\newpage

The $\nu$-process may contribute as well to production of some other
of the lowest abundance $p$-nuclei, like $^{11}$B and $^{19}$F  (Boyd 1999).
Generally, the process is thought to occur in the neutrino
wind generated by stellar collapse in supernovae. The nuclides
synthesized clearly depend on the shell in which the $\nu$-process occurs.
For example, $^{11}$B and $^{19}$F would be expected to be made
in shells in which the dominant constituents were $^{12}$C and $^{20}$Ne
respectively, both by
processes in which a neutrino would excite the target nucleus via
the neutral-current interaction (Boyd 1999). 

The $\nu$-process could also make two of the rarest stable
nuclides in the periodic table: $^{138}$La and $^{180}$Ta (Boyd 1999).
The latter would be made by the $^{181}$Ta($\nu,\rm{n})^{180}$Ta
(neutral current) reaction, which appears to produce an abundance consistent
with what observed. Similarly, the $^{139}$Ta($\nu,\rm{n})^{138}$La
(neutral current) reaction,
together with the
$^{138}$Ba($\nu,e)^{138}$La (charge current) reaction, appear capable
of synthesizing roughly the observed $^{138}$La abundance. Thus,
the $\nu$-process seems to provide a natural mechanism for synthesis of
$^{138}$La and $^{180}$Ta, which has evaded description for several decades, 
as well as some other nuclides (Boyd 1999). 
However, it should be noted that such results are somewhat uncertain due to
questions about the neutrino spectrum resulting from a Type II supernova
and many questions on neutrino processes have yet to be
solved in the future
(see, e.g., 
Woosley, Hartmann, R.D. Hoffman, \& Haxton 1990,
Boyd 1999,
Ginzburg 1999,
Henley \& Schifer 1999, 
Lang 1999,
Khlopov 1999,
Turner \& Tyson 1999,
Wolfenstein 1999, 
and references therein).	

\section{Spallation Cross Sections}
Precise nuclear spallation cross sections are needed in astrophysics not
only to calculate abundances of light elements with formulas of the
type (93) but also for many other tasks. So, it is believed today that
low energy cosmic ray interactions with ISM are responsible not only for
a part of LiBeB-production discussed in the previous section, but also 
for the
production of some of the now extinct radioisotopes that existed at the
time of the formation of the solar system
and found recently in meteorites, like $^{26}$Al, $^{41}$Ca,
and $^{53}$Mn (see Ramaty, Kozlovsky, \& Lingenfelter 1996a and
references therein). To estimate the abundances of these extinct 
radioisotopes
in the solar system one uses formulas similar to (93) and one needs
reliable cross sections for interaction of a variety of nuclei from the
LEC with H and $^4$He, the most abundant constituents of the
ambient medium (see details in
Ramaty, Kozlovsky, \& Lingenfelter 1996a).

Generally, a lot more spallation cross sections are needed to study 
meteorites besides the ones 
related with the extinct radioisotope production. During the recent
years, large number of meteorites were found on Antarctic icefields
and in hot deserts, in particular in the Sahara. These meteorite finds
have increased the interest in the investigation of cosmogenic nuclides.
Besides direct measurements of radionuclide composition performed for some
of the found meteorites, such investigations usually involve theoretical
calculations of production rates of cosmogenic nuclides in meteoroids
by folding depth- and size-dependent
spectra of primary and secondary cosmic-ray particles with the
cross sections of the underlying reactions. 
The quality and reliability of the calculated production rates
exclusively 
depend on the accuracy of the available spallation cross sections.
A serious progress
in interpretation the cosmogenic nuclide 
production in meteorites by galactic and solar cosmic rays
and in understanding the 
cosmic radiation itself
was achieved during the last years by the group of Prof. Rolf 
Michel at Hannover
(see, e.g., Michel, Leya, \& Borges 1996, Gilabert et al. 1998,
Weigel 1999,
and references therein). 

Another problem on cosmogenic nuclide production study requiring reliable
spallation cross sections from interactions of protons and
alphas up to
about 200 MeV with a variety of nuclei-targets is the investigation
of solar cosmic ray  (SCR) exposure of the lunar surface material,
as well as of the earth atmosphere (see, e.g., Bodemann et al. 1993
and references therein). The survey by Reedy and Marti (1990) may
serve as a good review on this subject and a source for further references.

As mentioned by Tsao, Barghouty, \& Silberberg (1999), it is believed
today that the elements Li, Be, B, Cl, K, Sc, Ti, V, Mn and much of N, 
Al, and P in cosmic rays (see Fig. 2) are produced by nuclear spallation
of the more abundant elements of the cosmic-ray source component,
i.e., C, O, Ne, Mg, Si, Ca, and Fe. Studies of the composition, propagation, 
and origin of galactic cosmic rays are still to a large degree model
dependent and conclusions made from such works depend essentially
on nuclear cross sections used in calculations, therefore
as precise as possible estimates of the relevant cross sections
are needed.

Let us mention just one more particular problem in astrophysics requiring
reliable cross sections. Recently, the gamma-ray line at 0.511 MeV
has been observed from a variety of astrophysical sites,
including solar flares
(see references in Kozlovsky, Lingenfelter, \& Ramaty 1987).
It is thought that this line is due to positron annihilation
on a electron ($e^+ + e^- \to \gamma + \gamma$, where one photon will
have a high energy and, if the electron is at rest, the other photon will 
have an energy on the order of $m_e c^2 = 0.511$ MeV)
from decay of radioactive nuclei and pions. One possibility of positron 
emitters production in a solar flare is via interactions of particles 
accelerated in the flare with the ambient solar atmosphere. To estimate
the annihilation of positrons from such radioactive nuclei one need 
to know a great variety of proton- and $\alpha$-induced spallation cross 
sections for the production of such positron emitters
(see details in Kozlovsky, Lingenfelter, \& Ramaty 1987).

The list of astrophysical tasks requiring reliable cross sections
can be continued further and further. As mentioned recently
by Waddington (1999), it appears that the most serious
limitation to deducting abundances of energetic nuclei in the
cosmic radiation arises not from our lack of astrophysical
measurements and observations, 
but just from our lack of the appropriate nuclear 
cross sections.
% (Waddington 1999).
Let us also note, that such spallation cross sections are 
of great
importance as well both for fundamental nuclear physics and for 
many nuclear applications, e.g., for
accelerator transmutation of waste (ATW), 
% for elimination of long-lived radioactive wastes with a spallation source,
accelerator-based conversion (ABC), 
% aimed to complete the destruction of weapon plutonium,
accelerator-driven energy production (ADEP), 
% which proposes to derive fission energy from thorium with 
% concurrent destruction of the long-lived waste
% and without the production of weapon-usable material, 
accelerator production of tritium (APT),
for the optimization of commercial production of radioisotopes  
used in medicine, mining, and industry, for solving problems of 
radiation protection of cosmonauts, aviators, workers at nuclear
facilities, and for modeling radiation damage to computer chips, etc.
(see details and references, e.g., in 
Mashnik, Sierk, Bersillon, \& Gabriel 1997).  
In the following subsections,
we present a short survey of available experimental,
calculated, and evaluated spallation cross sections  
for astrophysics and other fields together with our thought of 
how to possibly improve the present status of this problem.

%\vspace*{-20mm}
\subsection{Experimental Data}
Cosmic rays consist of all the elements in the periodic
table, up to uranium, therefore reactions induced by any projectile
are of interest for astrophysics. However,
since hydrogen is the dominant element, followed by helium,
spallation cross sections from
reactions induced by protons and alphas are of the first priority,
while we do mention as well the importance of 
nucleus-nucleus reactions for many astrophysics
problems, as surveyed recently by
Tsao, Barghouty, \& Silberberg (1999).
Thousands of measurements of 
spallation cross sections relevant 
to astrophysics 
(mainly, proton-induced)
have been performed over the last half a century.

A good survey of experimental cross sections for proton induced
spallation reactions measured before 1966,
was done by Bernas, Gradsztajn, Reeves, \& Schatzman (1967)
and included thereafter  
in Chapter 9 by Audoze, Epherre, \& Reeves (1967)
and Chapter 8 by Gradsztajn (1967)
of the well known book
{\em High-Energy Nuclear Reactions in Astrophysics} edited by B. S. P. Shen
and published by W. A. Benjamin, Inc. in 1967 in New York.
In a way, this survey
was like a ``bible" of nuclear cross sections in astrophysics,
as it was 
widely known and used, to our knowledge, without questions 
in almost all 
astrophysical simulations, up to very recent years.
A short but comprehensive review of experimental results
obtained by 1976 may be found in
Hudis (1976).
Another known in astrophysics paper serving as a
survey of both proton- and alpha-induced experimental
spallation cross sections was published 11 years later
(only figures, as a by-product)
by Kozlovsky, Lingenfelter, \& Ramaty (1987).
The last published short
astrophysical survey on spallation cross sections measurements
was, to our knowledge, the work by Tsao, Barghouty, \& Silberberg (1999).
Meanwhile, many other reliable measurements were performed
that are not covered by
these compilations, and, as one can see from Fig. 10,
not all old cross sections agree well with the new data.

Many efforts have been previously made as well by nuclear physicists
to compile 
experimental spallation cross sections from proton
and heavier projectiles induced reactions.
So, very good and 
comprehensive reviews of experimental excitation functions
from proton-, deuteron-, and alpha-induced reactions on a number
of light and medium nuclei-targets from Carbon to Chlorine, as well as
on Cu and Au, were published by Tobailem and co-authors from
1971 to 1983 at CEA, Saclay, France, in a convenient form of Reports
(in French) with tables and figures
(Tobailem et al. 1971, 1972, 1975, 1977,1981a, 1981b,
1982, and 1983).
But to the best of our knowledge,
the most complete compilation (ever published, in any fields
of nuclear cross sections data)
was performed by Sobolevsky and
co-authors at INR, Moscow, Russia, 
and was published by Springer-Verlag 
from 1991 to 1996 in eight
separate subvolumes 
(Sobolevsky et al. 1991, 1992, 1993, 1994a, 1994b,
1995, 1996a, and 1996b).
Sobolevsky and co-authors have performed a major work and compiled
all data available to them for target elements from Helium to 
transuranics for the entire energy range from thresholds up
to the highest energy measured.
For example,
for proton-induced reactions, this compilation contains about 37,000
data points  published in the first four Subvolumes, I/13a-d, 
(the following Subvolumes, I/13e-h, concern pion, antiproton,
deutron, triton, 
$^3$He,
and alpha induced reactions). 
This rich compilation is also currently available
in an electronic version as an IBM PC code named NUCLEX,
published only a month ago by 
Springer-Verlag 
in a hardcover format 

\eject

\begin{figure}[h!]
\vspace*{-1.0cm}
\hspace*{-12mm}
\psfig{figure=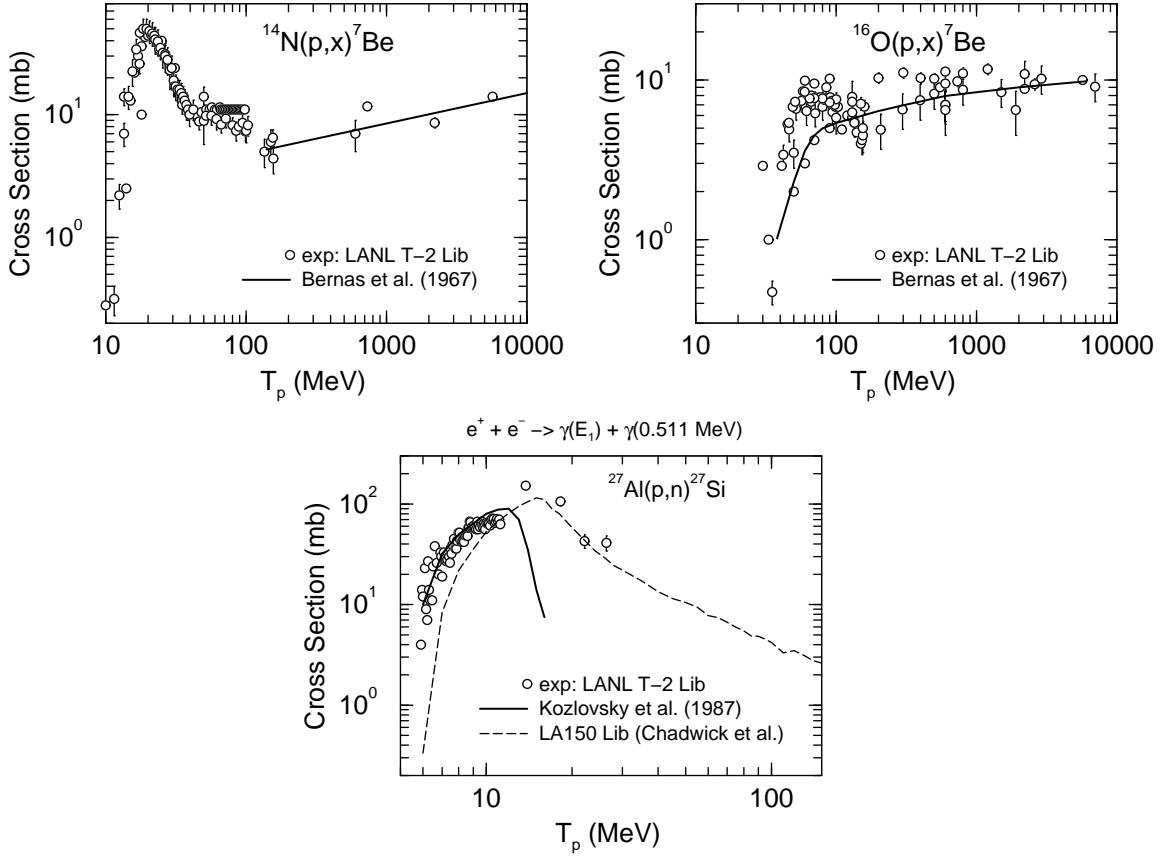,width=182mm,angle=0}
\vspace*{-10.0cm}

\caption{
Examples of cross sections used in astrophysical simulations
(solid lines) compared with the presently available data (open circles)
compiled in the LANL T-2 library (Mashnik et al. 1998).
On the two top plots, the cross sections for production of $^7$Be from
interaction of protons with $^{14}$N and $^{16}$O compiled in 1967 by
Bernas, Gradsztajn, Reeves, \& Schatzman and used thereafter in most
of the following
astrophysical simulations are shown together with 
available at present data (see text). 
The bottom plot shows the $^{27}$Al(p,n)$^{27}$Si cross section used by 
Kozlovsky, Lingenfelter \& Ramaty (1987) to evaluate the role of this
reaction as a positron-emitter in their interpretation
of the observed 0.511 MeV
line from solar flares. The 0.511 MeV line was interpreted in a model of
annihilation of positrons from the decay of radioactive nuclei produced
from interaction of particles accelerated in the flare with the 
ambient solar atmosphere.
For comparison, data from the 
LANL T-2 library are shown in this plot
together with the recent HMS-ALICE (Blann \& Chadwick 1998)
calculations by Chadwick (dashed line) 
from the LA150 transmutation/activation libraries (Koning, Chadwick,
MacFarlane, Mashnik, \& Wilson 1998).
}
\end{figure}

\newpage

{\noindent
accompanied by a CD-ROM with the NUCLEX code, 
as the ninth subvolume of this series and a suppliment to 
previous eight subvolumes
%\cite{nuclex}. 
(Sobolevsky et al. 2000;
see a detailed description of NUCLEX
in  Ivanov, Sobolevsky, \& Semenov (1998)).
}

Unfortunately this valuable compilation is either not known yet
by astrophysicists (we do not know any citations on it in
astrophysical papers) or is too expensive for individual users
and small 
libraries
(Springer Verlag
sells, e.g., the single subvolume I/13g
(Sobolevsky et al. 1996a)
for \$1647.00,
the subvolume I/13f,
(Sobolevsky et al. 1995)
for \$2020.00,
 and 
the last subvolume with the CD-ROM, I/13i
(Sobolevsky et al. 2000),
for \$2386.00; interested buyers 
may find information
on the Web, at pages: 
http://www.springer-ny.com/catalog/np/nov96np/DATA/3-540-61045-6.html, 
http://www.springer-ny.com/catalog/np/oct95np/DATA/3-540-59049-8.html, 
and, for the CD-ROM, on 
http://www.springer-ny.com/catalog/np/feb00np/3-540-63646-3.html).
%}
Of more immediate concern is the fact that NUCLEX does not contain 
a large volume of
data obtained during recent years, especially for proton-induced reactions. 

Due to the increasing interest in intermediate-energy data
for ATW, ABC, ADEP, APT, astrophysics, and other applications,  
precise and voluminous measurements of proton-induced spallation
cross sections have been performed recently, and are presently in progress,
by the group of Prof. Michel from Hannover University 
(see, 
e.g., 
Michel et. al. 1997,
Michel, Leya, \&~Borges~1996,~Gilabert~et~al. 
1998, and the Web
page \\ 
http://sun1.rrzn-user.uni-hannover.de/zsr/survey.htm\#url=overview.htm),
Yu. E. Titarenko et al. at ITEP, Moscow 
(Titarenko 1999a, 1999b, and references therein),
Yu. V. Aleksandrov et al. at JINR, Dubna 
(Aleksandrov et al. 1995 and references therein),
B. N. Belyaev et al. at B. P. Konstantinov St. Petersburg Institute of
Nuclear Physics 
(Belyaev, Domkin, \& Mukhin 
1994 and references therein),
N. I. Venikov et al. at Kurchatov Institute, Moscow 
(Venikov, Novikov, \& Sebiakin 1993),
A. S. Danagulyan et al. at JINR, Dubna 
(Danagulyan et al. 2000 and references therein),
H. Vonach et al. at LANL, Los Alamos 
(Vonach et al. 1997),
S. Sudar and S. M. Qaim at KFA, J\"ulich 
(Sudar \& Qaim 1994),
D. W. Bardayan et al. at LBNL, Berkeley 
(Bardayan et al. 1997),
J. M. Sisterson et al. at TRIUMF and other accelerators 
(Sisterson et al. 1997),
etc.
Finally, we note another, ``new" type of nuclear reaction
intensively studied in recent years, which provides irreplaceable
data both for nuclear astrophysics and nuclear physics itself.
These are from reactions using reverse kinematics,
when relativistic ions interact with hydrogen targets and they often
provide the only way to obtain reliable data for interaction
of intermediate energy 
protons with separate isotopes of an element with a complex
natural isotopic composition. Good data for this type of reactions
have been recently obtained, e.g.,
by W. R. Webber et al. at the LBL Bevalac 
(Webber, Kish, \& Schrier 1990, Chen 1997, and refences therein) 
and
L. Tassan-Got et al. at GSI, Darmstadt 
(Tassan-Got et al. 1998, Wlazlo et al. 2000).
Further references on several more such ``new" type of measurements,
as well as on recent spallation cross sections from nucleus-nucleus 
interactions may be found in
Silberberg, Tsao, \& Barghouty (1998) and
Tsao, Barghouty, \& Silberberg (1999).
These new data, as well as a number of other new and old measurements
have not been covered by NUCLEX. 

Let us note that for our needs, we compiled in the T-2 Group at LANL
also an experimental data library of spallation cross sections,
refered below as LANL T-2 Library (Mashnik, Sierk, Van Riper 
\& Wilson 1998). Our library is only for proton-induced reactions
and was completed so far only for 33 elements-targets:
C, N, O, F, Ne, Na, Mg, Al, P, S, Cl, Ar, K,
Ca, Fe, Co, Zn, Ga, Ge, As, Y, Zr, Nb, Mo, Sn, Xe, Cs, Ba, La, Ir,
Au, Hg, and Bi. But
 for the 91 targets (separate isotopes or natural
compostion) of these elements, our
library is the most complete, as far as we know,
and contains 23,439 data points covering 2,562 reactions, in comparison
with NUCLEX, having only 13,703 data points and 1594 reactions
for the same 33 elements. For these elements, we produced also
a calculated cross section library both for proton- and neutron-induced
reactions up to 5 GeV,
as well as an evaluated library,
discussed briefly in the next subsection. 

In developing our experimental LANL T-2 library,
we did not confine ourselves
solely to NUCLEX as a source of experimental cross sections; instead, we 
compile all available data for the targets in which we are interested,
searching first the World Wide Web, then any other sources available
to us, including the compilation from NUCLEX.
We also have begun to store in our library data for 
intermediate energy neutron-induced
reactions, but so far we have only 95 data points for Bi and C
targets covering 14 reactions induced by 
fast neutrons.
(Extensive neutron-induced experimental and evaluated
activation libraries at energies bellow 150 MeV
have been produced, validated, and used by many authors;
see, e.g., Muir \& Koning (1997), Korovin et al. (1999),
Chadwick et al. (1999), Fessler et al. (2000) and references
therein.)
Our library is still in progress, 
we permanently update it when new data for our elements
are available,
and we hope to extend
it, depending on our needs, and to make it available public through
the Web.

Note, that many 
data (especially, recent) on experimental spallation cross sections 
are
already included in the 
Experimental Nuclear Reaction Data Retrivals
(EXFOR)
compilation, available to users from
the Web through the international nuclear data banks
(see, e.g., the Web page of the NEA/OECD, Paris at
http://www.nea.fr/html/dbdata/dbexfor/html).  

From our point of view,
it would be useful for the astrophysical community to merge
the NUCLEX data library
(Sobolevsky et al. 1991-2000),
our
LANL T-2 compilation (Mashnik, Sierk, Van Riper, \& Wilson 1998),
and the data permanently updated in the EXFOR database
with 
already existing data libraries, considered 
by the Nuclear Astrophysics Data Effort Steering Committee
(Smith, Cecil, Firestone, Hale, Larson, \& Resler 1996)
as Nuclear Data
Resources for Nuclear Astrophysics, 
CSIRS (The Cross Section Information Storage and Retrieval System), 
ECSIL (The LLNL Experimental Cross Section Information Library),
and
ECSIL2 (a LANL/LLNL extension of ECSIL)
as well as to make available this information through the 
recent powerful NASA Astrophysical Data System
(Krutz, Eichhorn, Accomazzi, Grant, Murray, \& Watson 2000).
\subsection{Calculated and Evaluated Cross Sections}
Experiments to measure all data
necessary for astrophysics and other fields
are costly and there are a limited number of facilities 
available to make such measurements
(Blann et al. 1994, Nagel et al. 1995).
In addition, 
most measurements have been performed on targets with the natural composition of
isotopes for a given element and, what is more, often only cumulative
yields of residual product nuclei are measured. 
In contrast, 
for astrophysical simulations
and other applications, as well as 
to study the physics of nuclear reactions,
independent
yields obtained for isotopically separated targets are needed. 
Furthermore, only some 80--100 cross section values of residual product nuclei
are normally determined 
by the $\gamma$ spectrometry method
in the experiments with heavy nuclei, whereas, 
according to calculations, over 1000 residual product nuclei are 
actually produced.
Therefore, it turns out that reliable theoretical calculations are required 
to provide the necessary cross sections
(Blann et al. 1994, Nagel et al. 1995, Koning 1993).

In some cases, it is more convenient to have fast-computing 
semiempirical systematics for various characteristics
of nuclear reactions instead of using time-consuming, 
more sophisticated nuclear models. 
Therefore, to our knowledge,
in most astrophysical simulations one uses
predictions of different
semiempirical systematics 
(see, e.g., 
Silberberg, Tsao, \& Barghouty 1998,
Tsao, Barghouty, \& Silberberg 1999 and references therein).
After many years of effort by
many investigators, many empirical formulae are now available 
for spallation cross sections and excitation functions.
Many current systematics on excitation functions
have been reviewed by Koning (1993);
most of the old systematics available in 1970 were analyzed in the
comprehensive monograph
by Barashenkov and Toneev (1972);
the majority of systematics
for mass yields, charge dispersions, energy and angular
distributions of fragments produced in pA and AA collisions at
relativistic energies available in 1985
are presented in the review by H\"ufner (1985);
useful systematics for different hadron-nucleus interaction
cross sections may be found in our review
(Gabriel \& Mashnik 1996);
improved parametrizations for fragmentation cross sections
were recently published by S\"ummerer and Blank (2000); 
the last update of the well known and widely used 
in astrophysics code YIELD together with
further references may be found in
Silberberg, Tsao, \& Barghouty (1998) and
Tsao, Barghouty, \& Silberberg (1999).
Let us mentioned as well the old but widely used    
in the past in
astrophysical simulations
systematics by
Rudstam (1966), Gupta, Das, \& Biswas (1970), 
Silberberg \& Tsao (1973a, 1973b),
Foshina, Martins, \& Tavares (1984),
and direct 
readers interested in 
references on other phenomenological systematics to surveys 
by Koning (1993), Barashenkov \& Toneev (1972),
H\"ufner (1985),
Gabriel \& Mashnik (1996),
Tsao, Barghouty, \& Silberberg (1999),
as well as to the recent work by Michel et al. (1995).
Michel with co-authors (1995) have performed
a special analysis of predictabilities of different semiempirical
systematics and have concluded that ``Semiempirical
formulas will be quite successful if binding energies are the
crucial parameters dominating the production of the residual nuclides,
i.e. for nuclides far from stability. In the valley of stability,
the individual properties of the residual nuclei, such as level
densities and individual excited states, determine the final phase of 
the reactions. Thus, the averaging approach of all semiempirical formulas 
will be inadequate." 
In this case, one has to perform calculations in the 

\eject

\begin{figure}[h!]
\vspace*{-0.0cm}
\hspace*{-3mm}
\psfig{figure=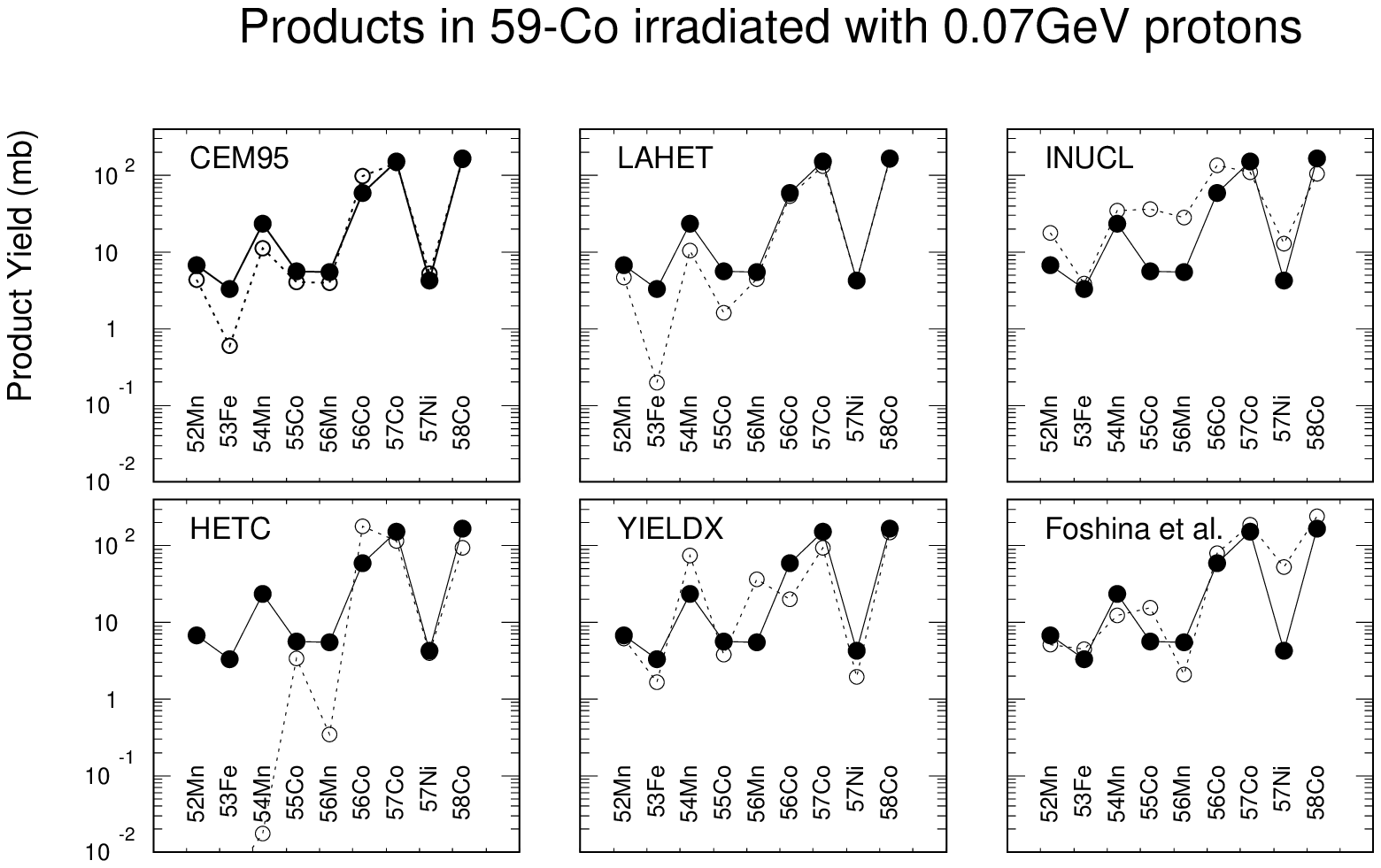,width=165mm,angle=0}
\vspace*{0.3cm}

\caption{
Product comparison between the new experimental (filled symbols) and
simulated (opaque symbols) yields in $^{59}$Co irradiated with
70-MeV protons (Titarenko et al. 1999a). Results labeled as YIELDX
and ``Foshina et al."
are obtained with the updated systematics by 
Silberberg, Tsao, \& Barghouty (1998) and using
the semiempirical formulas by  
Foshina, Martins, \& Tavares (1984), respectively,
that are often used in astrophysical simulations.
Results labeled as CEM95, LAHET, INUCL, and HETC were
calculated with Monte Carlo codes by
Mashnik (1995),
Prael \& Lichtenstein (1989),
Stepanov (1989),
and Armstrong \& Chandler (1972), respectively.
One can see discrepancies
more than an order of magnitude
for spallation cross sections of some isotopes. 
}
\end{figure}

{\noindent
framework of reliable
models of nuclear reactions. As was mentioned by Silberberg, Tsao, \&
Shapiro (1976),
there are also additional cases when 
Monte Carlo calculations should be used:
(1) when it is essential to know the distributions in angle and energy
for the ejected nucleons, (2) when the nuclear reaction is induced
by neutrons, and (3) when the particles have relatively low energies 
($E \le 60$ MeV).

As an example, Fig. 11 shows a comparison between the new data
for isotope production from interaction of 70-MeV
protons with $^{59}$Co
by Titarenko et al. (1999a) and 
results obtained 
with the 
systematics by
Silberberg, Tsao, \& Barghouty (1998), noted in figure as YIELDX, 
with semiempirical formulas by 
Foshina, Martins, \& Tavares (1984),
together with calculations using the Monte Carlo codes
CEM95 (Mashnik 1995),
LAHET (Prael \& Lichtenstein 1989),
INUCL (Stepanov 1989),
and HETC (Armstrong \& Chandler 1972).
One can see that for these reactions, neither
the phenomenological systematics by
Silberberg, Tsao, \& Barghouty (1998), 
nor the 
semiempirical formulas by 
Foshina, Martins, \& Tavares (1984),
both
widely used in astrophysics,
provide a good description of
all data, therefore we can not rely exclusively
on them in astrophysical and other simulations.
}

In such situations, 
one has to perform calculations in the framework of reliable
Monte Carlo
models of nuclear reactions and to use available experimental data.
As was mentioned 
by Mashnik, Sierk, Van Riper, \& Wilson (1998),
ideally, it would be desirable to have for applications
a universal evaluated library that includes data for 
all nuclides, projectiles, and incident energies. At present,
neither the measurements 
nor any
of the current models or phenomenological systematics
can be used alone to produce a reliable evaluated activation
library covering a large area of target nuclides and incident
energies. 
As one can see from Fig. 11, some of the
best Monte Carlo codes 
also have big difficulties in describing part of the data.
The problem is to find out the predictive power of 
different models, codes, and phenomenologilal systematics, and to
identify the regions of projectiles, targets, incident energies,
and produced nuclides where each model or systematics works better.
When we know this, we can create a reliable evaluated library
as we did in our medical isotope production study 
(Van Riper, Mashnik, \& Wilson 1998; 2000).
We think, a similar library
would be very useful
for astrophysical simulations as well, 
therefore let us remind here
our main consept.
We chose to create our evaluated library
(Mashnik, Sierk, Van Riper \& Wilson 1998)
by 
constructing excitation functions using all available
experimental data along with 
calculations 
using some of the more reliable codes, employing each of them in the 
regions of targets and incident energies where they are most applicable.
When we had reliable experimental data, they were taken as
the highest priority for our approximation as compared to model results,
and wherever possible, we attempted to construct a 
smooth transition from one data source to another.

The recent International Code Comparisons for Intermediate Energy 
Nuclear
Data organized by NEA/OECD at Paris 
(Blann et al. 1994, Michel \& Nagel 1997),
our own comprehensive benchmarks
(Van Riper et al. 1997,
Mashnik, Sierk, Van Riper \& Wilson 1998,
Van Riper, Mashnik, \& Wilson 1998 and 2000),
several studies by Titarenko et al. 
(1999a, 1999b, and refereces therein),
and
the recent Ph.D. thesis by Batyaev (1999),
specially dedicated to benchmark currently available
models and codes, 
have shown
that a modified version of the Cascade-Exciton model (CEM)
as realized in the code CEM95 
(Mashnik 1995)
and the LAHET code system 
(Prael \& Lichtenstein 1989)
generally have the best predictive powers for
spallation reactions at energies above 100 MeV as compared to other
available models. 

Therefore, we choose CEM95
(Mashnik 1995),
the recently improved version of the CEM code, CEM97x,
(Mashnik \& Sierk 1998),
and LAHET 
(Prael \& Lichtenstein 1989)
above 100 MeV to          
evaluate the required cross sections.
The same benchmarks have shown that at lower energies, the
HMS-ALICE code 
(Blann \& Chadwick 1998)
most accurately reproduces experimental results as compared with other 
models.
We therefore use the activation library calculated by Chadwick 
(M. B. Chadwick 1998, private communication)
with the HMS-ALICE code 
(Blann \& Chadwick 1998)
for protons
below 100 MeV and neutrons between 20 and 100 MeV. In the overlapping
region, between 100 and 150 MeV, we use both HMS-ALICE and CEM95 and/or
LAHET results. For neutrons below 20 MeV, we consider the data of the 
European
Activation File EAF-97, Rev. 1 
(Muir \& Koning 1996, Sublet, Kopecky, Forrest, \& Niegro 1997)
with some
recent improvements by
Herman (1996),
to be the most accurate results available;
therefore we use
them as the first priority in our evaluation.

Measured cross-section data from our LANL T-2 compilation described in 
the previus subsection
(Mashnik, Sierk, Van Riper \& Wilson 1998),
when available, are included together with theoretical results
and are used
to evaluate cross sections for study.
We note that when we put together all these different theoretical
results and experimental data, rarely do they agree perfectly with each
other, providing a smooth continuity of evaluated excitation functions.
Often, the resulting compilations show significant disagreement at energies
where the available data progresses from one source to another.
These sets are
thinned to eliminate discrepant data, providing data sets of more-or-less
reasonable continuity defining our evaluated cross sections. 

An examples with typical results of evaluated activation cross sections
for both proton- and neutron-induced reactions is shown in Fig. 12.
by broad gray lines. 51 similar color
figures for proton-induced reactions and 57 figures for neutrons,
can be found on the Web, in our detailed report 
(Van Riper, Mashnik, \& Wilson 1998).
We think that constructing and using similar evaluated libraries
in astrophysical calculations
(at least for the most important reactions)
would significantly improve the reliability of final results
and would help us, for instance, to better understand 
the origin of some light and medium 
elements, their abundances, and the role of
spallation processes in nucleosynthesis.

New reliable measurements, in particular, on
separate isotopes of (enriched) targets or
using reverse kinematics as mentioned above,
and further development of nuclear reaction
models and phenomenological systematics are necessary
to produce a reliable evaluated library of spallation cross sections.
Excitation functions, i.e., spallation cross sections 
as functions of the kinetic energy of projectiles,
are a very ``difficult" characteristic
of nuclear reactions as they involve together the different and complicated
physics processes of spallation, evaporation, fission, and fragmentation 
of nuclei. A lot of work is still necessary to be done by
theorists and code developers before a reliable complex of codes 
able to satisfactorily predict arbitrary excitation functions 
in a wide range of  
incident energies/projectiles/targets/final nuclides
will be available. At present, we are still very far from the completion 
of this difficult task (Mashnik, Sierk, Bersillon, \& Gabriel 1997).

In the meantime, to evaluate excitation functions
needed for astrophysics, nuclear
science, and applications, it is necessary to 
use and analyze together the available experimental data, and for each
region of incident energies/projectiles/targets/final nuclides,
the predictions of phenomenological systematics, and
the results of calculations with the most reliable codes, and not to
limit ourselves just to one source of data, as was practiced
in many past astrophysical simulations.

%\newpage

%\vspace*{-0.5cm}

\begin{figure}[t!]
\vspace*{-0.1cm}
\hspace*{5mm}
\psfig{figure=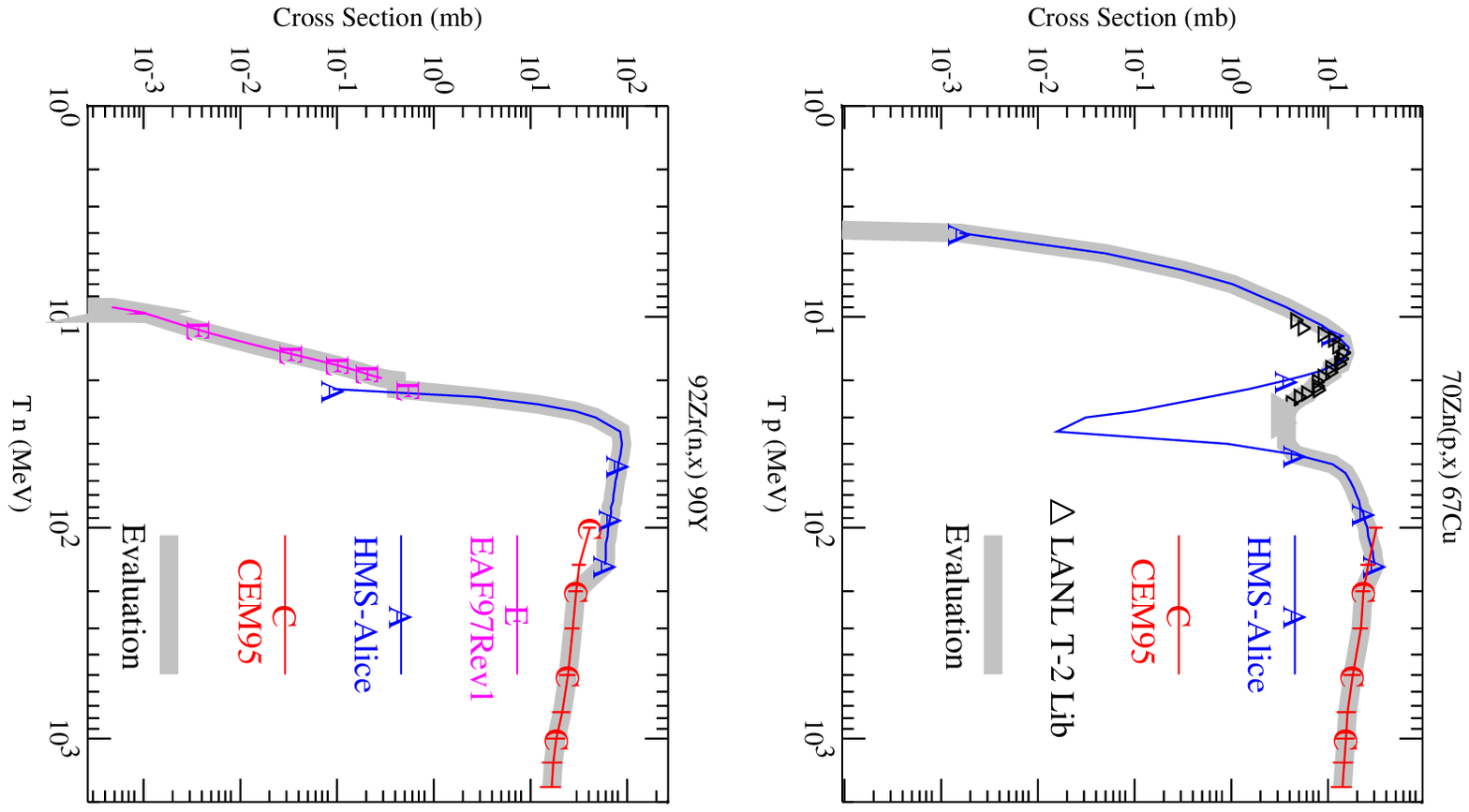,width=80mm,angle=270}
\end{figure}

\vspace*{8.2cm}

%\caption{
{\noindent Fig. 12.---
Examples of data and evaluations for (p,x) and (n,x) reactions from the
LANL T-2 library (Garland, Schenter, Talbert, Mashnik, \& Wilson 1999).
Experimental data for protons from 
the LANL T-2 compilation (Mashnik,
Sierk, Van Riper, \& Wilson 1998)
are shown by triangles, and for neutrons, from the European Activation
File EAF-97
(Sublet, Kopecky, Forrest, \& Niegro 1997),
by the magenta line marked with ``E". Calculations with the HMS-ALICE
code (Blann \& Chadwick 1998) are shown by blue lines marked with ``A",
and with the CEM95 code (Mashnik 1995), by red lines marked
with ``C".
Evaluated cross sections are shown by broad gray lines.
}
%\end{figure}

\newpage

\section{Summary}

We have performed
a brief review of nuclide abundances in the solar system and in cosmic rays 
and of the believed today mechanisms of their production.
We have shown on a number of examples that
nuclear spallation processes play an important role in synthesis 
not only of the light nuclei, Li-Be-B, but also
in production of other elements in the solar system,
in cosmogenic nucleosynthesis, 
in production of most energetic nuclei in cosmic rays,
in cosmic ray exposure of 
the lunar (and planets) surface
material and of meteorites, 
as a source of positron emitters, etc.

To study and understand these processes, reliable spallation
cross sections for a variety of reactions are needed.
We have performed a brief review of recent measurements,
compilations, calculations, and evaluations of spallation
cross sections relevant to astrophysics.
We have shown on several examples that in some past astrophysical
simulations old experimental cross sections were used
that are in poor agreement with recent measurements and
calculations with reliable modern models of nuclear reactions.

We suggest to not limit in astrophysical calculations only to
one source of spallation cross sections
as was done in some previous works
but to use 
and analyze together all available 
experimental data, and for each
region of incident energies/projectiles/targets/final nuclides,
the predictions of phenomenological systematics, and
the results of calculations with the most reliable models and codes.
Even better it would be to produce an universal evaluated 
library of spallation cross sections needed for astrophysics,
using together available experimental data and calculations
with the most reliable codes, as was done before in the
group T-2 at LANL for a number
of reactions of interest for our medical isotope production study.
Such an evaluated data library would be very useful not
only for astrophysical simulations, but also for fundamental nuclear
physics itself and a number of important applications, like
ATW, ABC, ADEP, APT, medical isotope production, etc.
New reliable measurements on
separate isotopes of (enriched) targets or
using reverse kinematics, extending and updating 
already created compilations of spallation cross sections
by nuclear physicists, like NUCLEX and the LANL T-2 library, as well as
merging these data libraries with astrophysical libraries,
like CSIRS, ECSIL, and ECSIL2, and,
finally, 
further development of nuclear reaction
models and phenomenological systematics are necessary
to successfully complete this goal. 

\acknowledgments

The author expresses his gratitude to Anna Hayes for initiating
this work, fruitful interactions, and help.
He is grateful to Karl-Ludwig Kratz and Gerhard Schmidt for
helpful discussions
and to Gerald Hale for careful
reading the draft of the paper and useful suggestions.
He thanks Robert MacFarlane, Daniel Strottman, and Laurie Waters for
stimulating interest and support, and acknowledges
Elisabeth Vangion-Flam,
Brian Fields, Keith Olive, Sean Ryan, Keneth Nollett, and Michael Turner
for  kind permisssion to reporoduce some of their 
figures in the present report.
This study was supported by the U. S. Department of Energy.
%and by ... (Anna, please add here your source of money spent
%on this work).

%\eject


\begin{thebibliography}{}

\baselineskip 0.15truecm

\bibitem{aleksandrov96}
%Yu.~V.~ALEKSANDROV, S.~K.~VASILJEV, R.~B.~IVANOV, L.~M.~KRIZHANSKY,
%M.~A.~MIKHAILOVA, T.~I.~POPOVA, V.~P.~PRIKHODTZEVA, V.~P.~EISMONT,
%A.~F.~NOVGORODOV, and R.~MISIAK,
%``Cross Section for Spallation Production of Radioactive
%Nuclides by 660 MeV Protons," 
%{\em Izv.~Rossiiskoi Akad.~Nauk, ser.~fiz.}, {\bf 59} (1995) 206
%  No.5, pp.206--210
%[{\em Bull.~Russian Acad.~Sci.: Physics}, {\bf 59} (1995) 895]
%and references therein.
% No.~5, pp.~895--899
Aleksandrov, Yu. V., et al. 1995,
Bull. Russian Acad. Sci.: Physics,  59, 895

\bibitem{angulo99}                  % The last compilation of Reaction Rates
%C. Angulo, M. Arnould, M. Rayet, P. Descouvemont, D. Baye, 
%C. Leclercq-Willain, A. Coc, S. Barhoumi, P. Aguer, C. Rolfs,
%R. Kunz, J. W. Hammer, A. Mayer, T. Paradellis, S. Kossionides,
%C. Chronidou, K. Spyrou, S. Degl`Innocenti, G. Fiorentini,
%B. Ricci, S. Zavatarelli, C. Providencia, H. Wolters, J. Soares,
% C. Grama, J. Rahighi, A. Shotter, M. Lamehi Rachti,
%``A Compilation of Charged-Particle Induced Thermonuclear Reaction Rates,"
% Nucl. Phys., {\bf A656} (1999) 3-183.
Angulo, C., et al. 1999,
Nucl. Phys. A., 656, 3

\bibitem{hetc}
%T.~W.~ARMSTRONG and K.~C.~CHANDLER,
%``HETC --- A High Energy Transport Code,"
%{\em Nucl.~Sci.~Eng.}, {\bf 49}, 110 (1972).
Armstrong, T. W., \& Chandler, K. C. 1972,
Nucl. Sci. Eng.,  49, 110 

\bibitem{arnett68}
% W. D. Arnett, ``On Supernova Hydrodynamics,"
% Astrophys. Journ., {\bf 153} (1968) 341.
Arnett, W. D. 1968,
ApJ, 153, 341

\bibitem{arnett69}
% W. D. Arnett, ``A Possible Model of Supernovae: Detonation of $^{12}$C,"
% Astrophys. and Space Sci. {\bf 5} (1969) 180.
Arnett, W. D. 1969,
%Astrothys. and Space Sci. 5, 180
\apss, 5, 180

\bibitem{arnett69a}
% W. D. Arnett, ``Explosive Nucleosynthesis in Stars,"
% Astrophys. Journ., {\bf 137} (1969) 1369.
Arnett, W. D. 1969,
%ApJ, 137, 1369
\apj, 137, 1369

\bibitem{arnett95}
% W. D. Arnett, ``Explosive Nucleosynthesis Revisited: Yields,"
% Ann. Rev. Astron. Astrophys., {\bf 33} (1995) 115. 
Arnett, W. D. 1995,
%Ann. Rev. Astron. Astrophys., 33, 115 
\araa, 33, 115

\bibitem{audouze67}
%J. Audouze, M. Epherre, and Hubert Reeves,
%``Survey of Experimental Cross Sections for Proton-Induced Spallation 
%Reactions in He$^4$, C$^{12}$, N$^{14}$, and O$^{16}$,"
%in: {\em High-Energy Nuclear Reactions in Astrophysics},
%ed. B. Shen, New York: W. A. Benjamin (1967), pp. 255-271.
Audouze, J., Epherre,  M., \& Reeves, H. 1967,
in: High-Energy Nuclear Reactions in Astrophysics,
ed. B. Shen (New York: W. A. Benjamin), 255

\bibitem{book}  
Barashenkov, V. S., \& Toneev, V. D. 1972,
Interaction of High Energy Particles
and Nuclei with Atomic Nuclei (Moscow: Atomizdat) 

\bibitem{bardayan97}  
%D. W. Bardayan, 
%M. T. F. da Cruz, M. M. Hindi, A. F. Barghouty, Y. D. Chan, 
%A. Garcia, R.-M. Larimer, K. T. Lesko, E. B. Norman, D. F. Rossi, 
%F. E. Wietfeldt, and I. Zlimen, 
%      "Radioactive Yields from 1.85-GeV Protons on Mo and 1.85- and 5.0-GeV 
% Protons on Te,"
%{\em Phys. Rev.}, {\bf C55} (1997) 820. 
Bardayan, D. W., et al. 1997, 
Phys. Rev., C55, 820 

\bibitem{belyaev94}  
%B. N. Belyaev, V. D. Domkin, and V. S. Mukhin, 
%``Isotopic Effects in Spallation and Fission Reactions and 
%Mass-Spectrometer Investigation of Them with 1-GeV Protons," 
%{\em Fiz. Elem. Chastits At. Yadra}, {\bf 23} (1992) 993
%[{\em Sov. J. Part. Nucl}., {\bf 23} (1992) 439];
%``Isotope Effects in Fragment Yields in Uranium Nuclear Fission 
%Induced by Intermediate-Energy Protons,
%{\em Yad. Fiz.}, {\bf 57} (1994) 1231 
%[{\em Phys. At. Nucl.}, {\bf 57} (1994) 1163].
Belyaev, B. N., Domkin,  V. D., \& Mukhin, V. S. 1994,
Phys. At. Nucl., 57, 1163

\bibitem{bernas67}      % Exp. data 14N(p,x)Be7 & 16O(p,x)Be7 shown in Fig.4
% R. Bernas, E. Gradsztajn, H. Reeves, and E. Schtzman,
%``On the Nucleosynthesis of Lithium, Berilium, and Boron,"
%Ann. Phys. (N.Y.), {\bf 44} (1967) 426-478.
Bernas, R., Gradsztajn, E., Reeves, H., \& Schtzman, E. 1967,
Ann. Phys. (N.Y.), 44, 426

\bibitem{bethe99}
% Hans A. Bethe,
%``Nuclear Physics,"
%Rev. Mod. Phys., {\bf 71}, S6 (1999).
Bethe, H. A. 1999,
Rev. Mod. Phys., 71, S6

\bibitem{nea94a}
%M.~BLANN, H.~GRUPPELAR, P.~NAGEL, and J.~RODENS, 
%{\em International Code Comparison for
%Intermediate Energy Nuclear Data},
%NEA OECD, Paris (1994).
Blann, M., Gruppelar, H., Nagel, P., \& Rodens, J. 1994,
International Code Comparison for
Intermediate Energy Nuclear Data,
(Paris: NEA OECD) 

\bibitem{hmsalice}
Blann, M., \& Chadwick, M. B. 1998,
Phys. Rev. C, 57, 233 


\bibitem{bodemann93}
%R. Bodemann, H.-J. Lange, I. Leya, R. Michel, 
%T. Schiekel, R. Rosel, U. Herpers,
%H. J. Hofmann, B. Dittrich, M. Suter, W. Wolfli, B. Holmqvist, H. Conde,
% and P. Malmborg,
%``Production of Residual Nuclei by proton-Induced Reactions on C, N, O, Mg,
%Al, and Si,"
%Nucl. Instr. Meth., {\bf B 82} (1993) 9-31.
Bodemann, R., et al. 1993,
Nucl. Instr. Meth. B, 82, 9

\bibitem{boyd99}
%Richard N. Boyd,
%``Nuclear Astrophysics at the Beginning of the Twenty First Century,"
%Chapter 22 in {\em Heavy Elements and Related New Phenomena, Vol. 2},
%edited by Walter Greiner and Raj K. Gupta, World Scientific,
%Singapore, 1999, pp. 893-975.
Boyd, R. N. 1999,
Chapter 22 in 
Heavy Elements and Related New Phenomena, Vol. 2,
eds. W. Greiner and R. K. Gupta (Singapore: World Scientific), 893

\bibitem{abible}
%E. M. Burbidge, G. R. Burbidge, W. A. Fowler, and F. Hoyle,
%``Synthesis of the Elements in Stars,"
%Rev. Mod. Phys., {\bf 29}, 547 (1957).
Burbidge, E. M., Burbidge, G. R., Fowler, W. A., \& Hoyle, F. 1957,
Rev. Mod. Phys., 29, 547 

\bibitem{batyaev99}
Batyaev, V. F. 1999, Ph.D. thesis, ITEP, Moscow

\bibitem{burles98a}
% Scott Burles and David Tytler,
%``The Deuterium Abundance Toward Q1937-1009,"
% Astrophys. J., {\bf 499}, 699-712 (1998).
Burles, S., \& Tytler, D. 1998a,
ApJ, 499, 699

\bibitem{burles98b}
% Scott Burles and David Tytler,
%``The Deuterium Abundance Toward QSO 1009+2956,"
% Astrophys. J., {\bf 507}, 732-744 (1998).
Burles, S., \& Tytler, D. 1998b,
ApJ, 507, 732

\bibitem{burles99}
% Scott Burles, Kenneth M. Nollett, James W. Truran, and Michael S. Turner,
%``Sharpening the Predictions of Big-Bang Nucleosynthesis,"
% Phys. Rev. Lett. {\bf 82}, No. 21, 4176-4179 (1999).
Burles, S., Nollett, K. M., Truran, J. W., \& Turner, M. S. 1999,
Phys. Rev. Lett., 82, 4176

%          A review on Nucleosynthesis in Asymptotic Giant Branch Stars (AGB)
\bibitem{busso99}
%M. Busso, R. Gallino, and G. J. Wasserburg,
%``Nucleosynthesis in Asymptotic Giant Branch Stars: Relevance for Galactic
% Enrichment and Solar System Formation,"
%Annu. Rev. Astron. Astrophys., {\bf 37} (1999) 239-309.
Busso, M., Gallino, R., \& Wasserburg, G. J. 1999,
ARA\&A 37, 239

%                                  An autobiographical review by Cameron
\bibitem{cameron99}
%A. G. W. Cameron,
%``Adventures in Cosmology,"
%Annu. Rev. Astron. Astrophys., {\bf 37} (1999) 1-36.
Cameron, A. G. W. 1999,
ARA\&A, 37, 1

%                                                             BBN
\bibitem{caso99}
% C. Caso et al., European Physical Journal, {\bf C3}, 1 (1998);
% Revised September 1999 by K. A. Olive, ``16. Big-Bang Nucleosynthesis,"
% available on the PDG WWW page: http://pdg.lbl.gov.
Caso, C, et al. 1998,
European Physical Journal C3, 1; 
Revised September 1999 update 
``16. Big-Bang Nucleosynthesis,"
by K. A. Olive, 
available on the PDG WWW page: http://pdg.lbl.gov

\bibitem{chadwick99}
%M. B. Chadwick, P. G. Young, S. Chiba, S. C. Frankle, G. M. Hale,
%H. G. Huges, A. J. Koning, R. C. Little, R. E. MacFarlane, R. E. Prael, 
%and L. S. Waters,
%``Cross-Sections to 150 MeV for Accelerator-Driven Systems and 
%Implimentation in MCNPX,"
%Nucl. Sci. Eng., {\bf 131} (1999) 293-328.
Chadwick, M. B., et al. 1999,
Nucl. Sci. Eng., 131, 293

\bibitem{chen97}
%C.-Z. Chen et al.,
%, S. Albergo, Z. Caccia et al., 
%"Systematics of Isotopic 
%       Production Cross Sections from Interactions of Relativistic Ca-40
%       in Hydrogen," 
%{\em Phys. Rev.}, {\bf C56} (1997) 1536.
Chen, C.-Z.,  et al. 1997,
Phys. Rev. C, 56, 1536

\bibitem{copi97}
% Craig J. Copi, David N. Schramm, and Michael S. Turner,
%``Big-Bang Nucleosynthesis Limit to the Number of Neutrino Species,"
% Phys. Rev. {\bf C33}, No. 6, 3389-3393 (1997).
Copi, C. J., Schramm, D. N., \& Turner, M. S. 1997,
Phys. Rev. C, 55, 3389 

%                                                             r-processes
\bibitem{cowan99}
% John J. Cowan, B. Pfeiffer, K.-L. Kratz, F.-K. Thielemann, 
% Christopher Sneden, Scott Burles, David Tytler, and Timothy C. Beers,
%``r-Process Abundaces and Chronometers in Metal-Poor Stars,"
% Astrophys. J., {\bf 521} (1999) 194-205.
Cowan, J. J., Pfeiffer, B., Kratz, K.-L., Thielemann, F.-K., 
Sneden, C., Burles, S., Tytler, D., \& Beers, T. C. 1999,
ApJ, 521, 194

\bibitem{crosas96}                                       % to cite: Director
%M. Crosas and J. Weisheit,
%``Spallation in Active Galactic Nuclei,"
%Astrophys. J., {\bf 465} (1996) 659-665.  
Crosas, M., \& Weisheit, J. 1996,
ApJ, 465, 659  

\bibitem{danagulyan00}
%A. S. Danagulyan, J. Adam, A. R. Balabekyan, V. G. Kalinnikov,
%V. I. Stegailov, V. K. Rodionov, V. I. Fominikh, and J. Frana,
%``Production of Light Nuclei an the Reactions of Protons with
%Separated Tin Isotopes,"
%Yad. Fiz., {\bf 63}, No. 2 (2000) 204-208
%[Phys. At. Nucl., {\bf 63}, No. 2 (2000) 151-155].
Danagulyan, A. S., et al. 2000,
Phys. At. Nucl., 63, 151

\bibitem{duncan92}
%Douglas K. Duncan, David L. Lambert, and Michael Lemke,
%``The Abundance of Boron in Three Halo Stars,"
%Astrophys. J., {\bf 401} (1992) 584-592.  
Duncan, D. K., Lambert, D. L., \& Lemke, M. 1992,
ApJ, 401, 584

%Importance of spallation processes in nucleosynthesis outside the Solar system
\bibitem{duncan97}                   
%D. K. Duncan, F. Primas, L. M. Rebull, A. M. Boesgaard,
% Constantine P. Deliyannis, L. M. Hobbs, J. R. King, and S. G. Ryan,
%``The Evolution of Galactic Boron and the Production Site of the Light
% Elements," Astrophys. J., {\bf 488} (1997) 338-349.  
Duncan, D. K., et al. 1997,
ApJ, 488, 338

\bibitem{fessler00}
%A. Fessler, A. J. M. Plompen, D. L. Smith, J. W. Meadows, and Y. Ikeda,
%``Neutron Activation Cross-Section Measurements from 16 to 20 MeV
%for Isotopes F, Na, Mg, Al, Si, P, Cl, Ti, V, Mn, Fe, Nb, Sn, and Ba,"
%Nucl. Sci. Eng., {\bf 134} (2000) 171-200                   
Fessler, A., et al. 2000,
Nucl. Sci. Eng., 134, 171

\bibitem{fieldsolive98}                   
% Brian D. Fields and Keith A. Olive,
%``The Evolution of $^6$Li in Standard Cosmic-Ray Nucleosynthesis,"
% Preprint UMN-TH-1727/98; TPI-MINN-98/22; Eprint astro-ph/9811183.
Fields, B. D., \& Olive, K. A. 1998,
preprint (astro-ph/9811183)

\bibitem{fields99a}                   
% Brian D. Fields, Keith A. Olive, Elisabeth Vangioni-Flam,
% and Michael Casse,
%``Testing Spallation Processes with Berylium and Boron," 
% Preprint UMN-TH-1826/99; TPI-MINN-99-51; Eprint astro-ph/9911320.
Fields, B. D., Olive, K. A., Vangioni-Flam, E., \& Casse, M. 1999,
preprint (astro-ph/9911320)

% A one -zone (closed box) GCE Model used as an example of neutrino synthesis
\bibitem{fields99b}                   
% Brian D. Fields and Keith A. Olive,
%``The Revival of Galactic Cosmic Ray Nucleosynthesis?" 
% Preprint UMN-TH-1719/98; Eprint astro-ph/9809277 v2;
%Astrophys. J., {\bf 516}, in press (1999).
Fields, B. D., \& Olive, K. A. 1999,
ApJ, 516, in press (astro-ph/980977 v2)

\bibitem{fo84}
%M.~FOSHINA, J.~B.~MARTINS, and O.~A.~P.~TAVARES,
%``Systematics of Spallation Yields with a Four-Parameter Formula,"
%{\em Radiochim.~Acta}, {\bf 35}, 121 (1984).
% No.~3, p.~121--131.
Foshina, M., Martins,  J. B., \& Tavares,  O. A. P. 1984,
Radiochim. Acta, 35, 121

\bibitem{systrev96}
%T.~A.~GABRIEL and S.~G.~MASHNIK,
%``Semiempirical Systematics for Different Hadron-Nucleus
%Interaction Cross Sections,"
%JINR Preprint E4-96-43, Dubna (1996).
Gabriel, T. A., \& Mashnik, S. G. 1996,
JINR Preprint E4-96-43, Dubna

\bibitem{garland99}
%M. A. Garland, R. E. Schenter, R. J. Talbert, S. G. Mashnik, 
%and W. B. Wilson, ``Nuclear Data Requirement for the Production of
%Medical Isotopes in Fission Reactors and Particle Accelerators,"
%Proc. 3rd Int. Conf. on Isotopes, Vancouver, Canada, September 6-10, 1999;
%LANL Report LA-UR-99-4898 and PNNL-SA-32104, 1999;
%Eprint: {\bf physics/9909021}.
Garland, M. A., Schenter,  R. E., Talbert, R. J., Mashnik, S. G., 
\& Wilson, W. B. 1999,
Proc. 3rd Int. Conf. on Isotopes, Vancouver, Canada, September 6-10, 1999,
in press, preprint LA-UR-99-4898 (physics/9909021)

\bibitem{gilabert98}
%E. Gilabert, B. Lavielle, S. Neumann, M. Gloris, R. Michel, Th. Schiekel,
%F. Sudbrock, and U. Herpes,
%``Cross Sections for the Proton-Induced Production of Krypton Isotopes
%from Rb, Sr, Y, and Zr for Energies up to 1600 MeV,"
%Nucl. Instr. Meth., {\bf B145} (1998) 293-319.
Gilabert, E., et al. 1998,
Nucl. Instr. Meth. B, 145, 293

\bibitem{ginzburg99}
% V. L. Ginzburg,
%``What Problems of Physics and Astrophysics Seem Now to be Especially
%Important and Interesting (Thirty Years Late, Already on the Verge of XXI
% Century,"
%Uspekhi Fizicheskikh Nauk, {\bf 169} No. 4, 419-441 (1999)
%[Physics-Uspekhi, {\bf 42} (4) 353-373 (1999)].
Ginzburg, V. L. 1999,
Physics-Uspekhi, 42, 353

\bibitem{gradsztajn67}
%Elie Gradsztajn,
%``Production of Li, Be, and B Isotopes in C, N, and O," 
%in: {\em High-Energy Nuclear Reactions in Astrophysics},
%ed. B. Shen, New York: W. A. Benjamin (1967), pp. 247-254.
Gradsztajn, E. 1967,
in: High-Energy Nuclear Reactions in Astrophysics,
ed. B. Shen (New York: W. A. Benjamin), 247

\bibitem{gupta70}
%B.~K.~GUPTA, S.~DAS, and M.~M.~BISWAS,
%``Cross Sections for the Production of Nuclides from Medium-Weight
%Elements by High-Energy Proton Bombardment,"
%{\em Nucl.~Phys.~A}, {\bf 155}, 49 (1970).
% No.1, p.~49--69.
Gupta, B. K., Das, S., \& Biswas, M. M. 1970,
Nucl. Phys. A, 155, 49

%                 A review on Quasistelar Object [Qasar (OSO)] Abundabces
\bibitem{hamann99}
%F. Hamann and G. Ferland,
%``Elemental Abundances in Quasistelar Objects: Star Formation and Galactic
% Evolution at High Redshifts,"
%Annu. Rev. Astron. Astrophys., {\bf 37} (1999) 487-531.
Hamann, F. \& Ferland, G. 1999,
ARA\&A 37, 487

\bibitem{HenleySchiffer99}
%E. M. Henley and J. P. Schiffer,
%``Nuclear Physics at the End of the Century,"
%Rev. Mod. Phys., {\bf 71}, No. 2, S205-S219 (1999).
Henley, E. M., \& Schiffer, J. P. 1999,
Rev. Mod. Phys., 71, S205

\bibitem{herman96}
Herman, M. 1996,
LANL Report LA-UR-96-4914

\bibitem{hoyle46}
%F. Hoyle,
%``The Synthesis of the Elements from Hydrogen,"
%Mon. Not. R. Astron. Soc. {\bf 106}, 343 (1946).
Hoyle, F. 1946,
Mon. Not. R. Astron. Soc., 106, 343

\bibitem{hudis76} 
%J.~HUDIS,       	
%``Spallation Reactions: Experimental Results, 1967--1975,"
%pp.~9--25 in {\em Spallation Nuclear Reactions and Their Applications},
%{\bf Astrophysics and Space Science Library, vol.~59}, B.~S.~P.~SHEN and
%M.~MERKER, Eds., D.~Reidel Publishing Company, Dordrecht-Holland (1976).
Hudis, J. 1976,
in Spallation Nuclear Reactions and Their Applications,
Astrophysics and Space Science Library, vol.~59, ed.
B. S. P. Shen and M. Merker, 
(Dordrecht-Holland: D.~Reidel Pub. Comp.), 9

\bibitem{hufner85}
%J.~H\"UFNER,
%``Heavy Fragments Produced in Proton-Nucleus and Nucleus-Nucleus
%Collisions at Relativistic Energies,"
%{\em Phys.~Rep.}, {\bf 125}, 129 (1985).
H\"ufner, J. 1985,
Phys. Rep., 125, 129

\bibitem{ivanov98} 
%V.~I.~Ivanov, N.	~M.~Sobolevsky, and V.~G.~Semenov,
%``Computer Version Of the Handbook on Radionuclide Production
%Cross-Sections at Intermediate Energies (the NUCLEX Code),"
%{\em Proc. 3d Specialists Meeting on Shielding Aspects of Accelerators,
%Targets and Irradiation Facilities (SATIF-3)}, Tohoku University,
%Sendai, Japan, May 12-13, 1997, NEA/OECD (1998) 
%p. 277.
Ivanov, V. I., Sobolevsky, N. M., \& Semenov, V. G. 1998,
in Proc. 3d Specialists Meeting on Shielding Aspects of Accelerators,
Targets and Irradiation Facilities (SATIF-3), Tohoku University,
Sendai, Japan, May 12-13, 1997, 
(Paris: NEA/OECD), 277

%                    Primordial nucleosynthesis of C and heavier elements
\bibitem{kajno93}
%T. Kajino,
%``Primordial Nucleosynthesis and Evolution of the Light Elements:
%Cosmology and Cosmic Rays,"
%Proc. Int. Symp. on Origin and Evolution of the Elements,"
% Tokyo, Japan, 16-17 October 1992, Eds. S. Kubono and T. Kajino,
% World Scientific, Singapore, 1993, pp. 15-33.
Kajino, T. 1993,
in
Proc. Int. Symp. on Origin and Evolution of the Elements,
Tokyo, Japan, 16-17 October 1992, 
Ed. S. Kubono \& T. Kajino (Singapore: World Scientific), 15

%                    A recent review on Experimental Nuclear Astrophysics
\bibitem{kappeler98}
%F. K\"appeler, F.-K. Thielemann, and M. Wiescher,
%``Current Quest in Nuclear Astrophysics and Experimental Approaches,"
%Annu. Rev. Nucl. Part. Sci., {\bf 48} (1998) 175-251.
K\"appeler, F., Thielemann, F.-K., \& Wiescher, M. 1998,
Annu. Rev. Nucl. Part. Sci., 48, 175

%                                     A goog recent review on s Process
\bibitem{kappeler99}
%F. K\"appeler,
%``The Origin of the Heavy Elements: The s Process,"
%Prog. Part. Nucl. Phys., {\bf 43} (1999) 419-483.
K\"appeler, F. 1999, Prog. Part. Nucl. Phys., 43, 419 

\bibitem{khlopov99}
% Maxim Yu. Khlopov,
%``Cosmoparticle Physics,"
% World Scientific Publishing Co. Pte. Ltd.,
%Singapore (1999).
Khlopov, M. Yu. 1999,
Cosmoparticle Physics,
(World Scientific; Singapore)

\bibitem{koning93}
%A.~J.~KONING,
%``Review of High Energy Data and Model Codes for Accelerator-Based
%Transmutation,"
%ECN-C-93-005, Petten (January 1993).
Koning, A. J. 1993,
ECN-C-93-005 Report, Petten

\bibitem{koning98}
%A. J. Koning, M. B. Chadwick, R. E. MacFarlane, S. G. Mashnik, and 
% W. B. Wilson,
%``Neutron and Proton Transmutation-Activation Cross Section Libraries
% to 150 MeV for Applications in Accelerator-Driven Systems and 
%Radioactive Ion Beam Target-Design Studies,"
% ECN-R-98-012 Report, Petten (1998).
Koning, A. J., Chadwick, M. B., MacFarlane, R. E., Mashnik, S. G., 
\& Wilson, W. B. 1998,
ECN-R-98-012 Report, Petten

\bibitem{korovin99} 
%Yu. Korivin, A. Konobeyev, P. Pereslavtsev, A. Stankovsky,
%U. Fischer, and U. von M\"ollendorff,
%``Intermediate Energy Activation File (IEAF-99),"
%Proc. 3rd Int. Conf. on Accelerator-Driven Transmutation Technologies
%and Applications (ADTTA'99), Praha, Czech Republic, June 7-11, 1999
%(Paper \# P-C29 on the Web page http://www.fjfi.cvut.cz/con_adtt99/). 
Korovin, Yu., et al. 1999, in
Proc. 3rd Int. Conf. on Accelerator-Driven Transmutation Technologies
and Applications (ADTTA'99), Praha, Czech Republic, June 7-11, 1999,
(Paper \# P-C29 on the Web page http://www.fjfi.cvut.cz/con\_adtt99/)  

\bibitem{kozlovsky87}
%B. Kozlovsky, R. E. Lingenfelter, and R. Ramaty, 
%``Positrons From Accelerated Particle Interactions,"
%Astrophys. J., {\bf 316} (1987) 801-818.  
Kozlovsky, B., Lingenfelter, R. E., \& Ramaty, R. 1987,
ApJ, 316, 801

\bibitem{krauss90}   % The best of the old reviews on BBNucleosynthesis !!! 
%Lawrence M. Krauss and Paul Romanelli,
%``Big Bang Nucleosynthesis: Predictions and Uncertainties,"
%Astrophys. J., {\bf 358} (1990) 47-59.  
Krauss, L. M., \& Romanelli, P. 1990,
ApJ, 358, 47

\bibitem{kurtz00}
Kurtz, M. J., Eichhorn, G., Accomazzi, A., Grant, C., Murray, S. S.,
\& Watson, J. M. 2000,
The NASA Astrophysical Data System: Overview,
in press, (arXiv:astro-ph/0002104)


\bibitem{lang80}
Lang, K. R. 1980,
Astrophysical Formulae 
%(A Compendium for the Physicist and Astrophysicist), 
(2d ed.; Berlin: Springer-Verlag)

\bibitem{lang99}
Lang, K. R. 1999,
Astrophysical Formulae, 2 volumes
% I: Radiation, Gas Processes and High Energy Astrophysics
%) II: Space, Time, Matter and Cosmology 
%(A Compendium for the Physicist and Astrophysicist), 
(3d enlarged and revised ed.; Berlin: Springer-Verlag)

\bibitem{tobailem72} 
Lassus St-Genies, C. H., \& Tobailem, J. 1972,
Note CEA-N-1466(2), Saclay

\bibitem{cem95}
Mashnik, S. G. 1995,
User Manual for the Code CEM95, Joint Institute for
Nuclear Research, Dubna, Russia; see the Web page
http://www.nea.fr/abs/html/iaea1247.html

\bibitem{ourlib}
%S. G. Mashnik, A. J. Sierk, K. A. Van Riper, and W. B. Wilson,
%{\em ``Production and Validation of Isotope Production Cross Section 
%Libraries for Neutron and Protons to 1.7 GeV}," 
%Proc. Fourth Workshop on Simulating Accelerator Radiation Environments, 
%Sept. 14--16, 1998, Knoxville, Tenn, pp. 151--162; also available as 
%Eprint {\bf nucl-th/9812071} on the LANL xxx.lanl.gov server.
Mashnik, S. G., Sierk, A. J., Van Riper, K. A., \& Wilson, W. B. 1998,
in Proc. 4th Workshop on Simulating Accelerator Radiation Environments, 
Sept. 14-16, 1998, Knoxville, TN, 
ed. T. A. Gabriel (Oak Ridge: ORNL), 151

\bibitem{cem97x}
Mashnik, S. G., \& Sierk, A. J. 1998,
in Proc. 4th Workshop on Simulating Accelerator Radiation Environments, 
Sept. 14-16, 1998, Knoxville, TN, 
ed. T. A. Gabriel (Oak Ridge: ORNL), 29

\bibitem{report97}
%S. G. Mashnik, A. J. Sierk, O. Bersillon, and T. A. Gabriel,
%{\em ``Cascade-Exciton Model Detailed Analysis of Proton Spallation at
%Energies from 10 MeV to 5 GeV}," 
%LANL Report LA-UR-97-2905 (1997); 
%http://t2.lanl.gov/publications/publications.html.
Mashnik, S. G.,  Sierk, A. J., Bersillon, O., \& Gabriel, T. A. 1997,
Cascade-Exciton Model Detailed Analysis of Proton Spallation at
Energies from 10 MeV to 5 GeV, 
LANL Report LA-UR-97-2905, 
http://t2.lanl.gov/publications/publications.html.

%                 A review on Chemical Evolution and Abundabces in the Galaxy
\bibitem{mcwilliam97}
%A. McWilliam,
%``Abundance Ratios and Galactic Chemical Evolution,"
%Annu. Rev. Astron. Astrophys., {\bf 35} (1997) 503-556.
McWilliam, A. 1997,
ARA\&A 35, 503

\bibitem{michaud72}           % Quasiequilibrium processes
%G. Michaud and W. A. Fowler,
%``Nucleosynthesis in silicon burning,"
% Astrophys. Journ., {\bf 173} (1972) 157.
Michaud, G., \& Fowler, W. A.
1977, ApJ, 173, 157

\bibitem{michel97}
%R.~MICHEL, R.~BODEMANN, H.~BUSEMANN, R.~DAUNKE, M.~GLORIS, H.-J.~LANGE,
%B.~KLUG, A.~KRINS, I.~LEYA, M.~L\"UPKE, S.~NEUMANN, H.~REINHARDT,
%M.~SCHNATZ-B\"UTTGEN, U.~HERPERS, Th.~SCHIENKEL, F.~SUDBROCK, B.~HOLMQVIST,
%H.~COND\'E, P.~MALMBORG, M.~SUTER, B.~DITTRICH-HANNEN, P.-W.~KUBIK,
%H.-A.~SYNAL, and D.~FILGES
%``Cross Sections for the Production of Residual Nuclides by Low- and
%Medium-Energy Protons from the Target Elements C, N, O, Mg, Al, Si, Ca, Ti, V, 
%Mn, Fe, Co, Ni, Cu, Sr, Y, Zr, Nb, Ba, and Au,"
%{\em Nucl. Instr.~Meth.}, {\bf B129} (1997) 153;
%see also the Web page at:\\
%{\bf http://sun1.rrzn-user.uni-hannover.de/zsr/survey.htm\#url=overview.htm}.
Michel, R. et al. 1997,
Nucl. Instr. Meth. B, 129, 153

\bibitem{michelnagel97}
Michel, R., \&  Nagel, P. 1997,
International Codes and
Model Intercomparison for Intermediate Energy Activation Yields,
NSC/DOC(97)-1,
(Paris: OECD) 

\bibitem{michel96}
%R. Michel, I. Leya, L. Borges,
%``Production of Cosmogenic Nuclides in Meteoroids: Accelerator Experiments
%and Model Calculations to Desipher the Cosmic Ray Record
%in Extraterrestrial Matter,"
%Nucl. Instr. Meth., {\bf B113} (1996) 434-444.
Michel, R., Leya, I., \& Borges, L. 1996,
Nucl. Instr. Meth. B, 113, 434

\bibitem{michel95}
%R.~MICHEL, M.~GLORIS, H.-J.~LANGE, I.~LEYA, M.~L\"UPKE, U.~HERPERS,
%B.~DITTRICH-HANNEN, R.~R\"OSEL, Th.~SCHIEKEL, D.~FILGES, P.~
%DRAGOVITSCH, M.~SUTER, H.-J.~HOFMANN, W.~W\"OLFLI, P.-W.~KUBIK,
%H.~BAUR, and R.~WIELER,
%``Nuclide Production by Proton-Induced Reactions on Elements
%($6 \le Z \le 29$) in the Energy Range from 800 to 2600 MeV,"
%{\em Nucl.~Instr.~Meth.~B}, {\bf 93}, 183 (1995).
% No.~?, p.~183--222.~
Michel, R. et al. 1995,
Nucl. Instr. Meth. B, 93, 183

\bibitem{muirkoning96}
%D. W. Muir and A. J. Koning,             	
%``Validation of the ECNAF Neutron Activation Cross-Section Library,"
%Proc. 2d Int. Conf. on Accelerator-Driven Transmutation Technologies and
%Applications, Kalmar, Sweden, June 3-7, 1996, 
%ed., H. Cond\'e, (Stokholm: Gotab, 1997),
%pp. 469-475.
Muir, D. W., \& Koning, A. J. 1997,
in Proc. 2d Int. Conf. on Accelerator-Driven Transmutation Technologies and
Applications, Kalmar, Sweden, June 3-7, 1996, 
ed., H. Cond\'e, (Stokholm: Gotab), 469

\bibitem{nagel95}
%P.~NAGEL, J.~RODENS, M.~BLANN, and H.~GRUPPELAR,
%``Intermediate Energy Nuclear Reaction Code Intercomparison:
%Application to Transmutation of Long-Lived Reactor Wastes,"
%{\em Nucl.~Sci.~Eng.}, {\bf 119}, 97 (1995).
Nagel, P., Rodens, J., Blann, M., \& Gruppelar, H.,  1995,
Nucl. Sci. Eng., 119, 97

\bibitem{olive94}                                % Neutrino nucleosymthesis 
% Keith A. Olive, Nikolas Prantzos, Sean Scully, and Elizabeth Vangioni-Flam, 
%``Neutrino Process Nucleosynthesis and the $^{11}$B/$^{10}$B Ration,"
% Astrophys. J., {\bf 424} (1994) 666-670.
Olive, K. A., Prantzos, N., Scully, S., \& Vangioni-Flam, E.
1994, ApJ, 424, 666

\bibitem{olive99}                          % A recent good review, on BBN 
% K. A. Olive,
% ``Primordial Big Bang Nucleosynthesis"
% Preprint UMN-TH-1735/99, TPI-MINN-98/30, Eprint astro-th/99011231;
% Summary of Lectures given at the Advanced School on Cosmology and Particle
% Physics, Peniscola, Spain, June 1998 and at the Theoretical and
% Observational Cosmology Summer School, Cargese, Corsica, France, Aug. 1998.
Olive, K. A. 1999,
preprint UMN-TH-1735/99, TPI-MINN-98/30 (astro-th/99011231)

\bibitem{lahet}
%R.~E.~PRAEL and H.~LICHTENSTEIN,
%``User Guide to LCS: The LAHET Code System,"
%{\em LA-UR-89-3014}, Los Alamos National Laboratory (September 1989);
Prael, R. E., \& Lichtenstein, H. 1989,
LANL Report LA-UR-89-3014

\bibitem{ramaty96a}
%Reuven Ramaty, Benzion Kozlovsky, and Richard E. Lingenfelter, 
%``Light Isotopes, Extinct Radioisotopes, and Camma-Ray Lines from
%Low-Energy Cosmic-Ray Interactions,"
%Astrophys. J., {\bf 456} (1996) 525-540.  
Ramaty, R., Kozlovsky, B., \&  Lingenfelter, R. E. 1996a,
ApJ, 456, 525

\bibitem{ramaty96b}
%R. Ramaty,
%``Interstellar Gamma-Ray Lines From Low Energy Cosmic Ray Interactions,"
%Astronomy \& Astrophysics Supplement Series, {\bf 120} (1996) 373-380.  
Ramaty, R. 1996b, A\&AS, 120, 373

\bibitem{ramaty97}
%Reuven Ramaty, Benzion Kozlovsky, Richard E. Lingenfelter, and Hubert Reeves,
%``Light Elements and Coswmic Rays in the Early Galaxy,"
%Astrophys. J., {\bf 488} (1997) 730-748.  
Ramaty, R., Kozlovsky, B., Lingenfelter, R. E., \& Reeves, H. 1997,
ApJ, 488, 730

\bibitem{ramaty98}
%R. Ramaty, B. Kozlovsky, and R. Lingenfelter,
%``Cosmic Rays, Nuclear Gamma Rays and the Origin of the Light Elements,"
%Physics Today, April 1998, pp. 30-35.
Ramaty, R., Kozlovsky, B., \& Lingenfelter, R. E. 1998,
Phys. Today, 51, 4, 30

\bibitem{ramaty99}
%Reuven Ramaty, Sean T. Scully, Richard E. Lingenfelter, and Benzion Kozlovsky,
%``Light Element Evolution and Cosmic Ray Energetics,"
% Eprint astro-ph/9909021, submitted to the APJ.
Ramaty, R., Scully, S. T., Lingenfelter, R. E., \& Kozlovsky, B. 1999,
ApJ, in press (astro-ph/9909021)

\bibitem{LiBeB98}
Ramaty, R., Vangioni-Flam, E., Cass\'e, M., \& Olive, K. 1999,
eds., Proc. Conf. on LiBeB Cosmic Rays and Gamma-Ray Line
Astronomy, Paris, December 1998, ASP Conf. Series, 171
(San Francisco: ASP)

\bibitem{reehy90}
Reedy, R. C., \& Marti, K. 1990,
in The Sun in Time, eds. C. P. Sonett, M. S. Giampapa, \& M. S. Matthews
(Tucson: Univ. of Arizona Press), 260

\bibitem{reeves94}
%Hubert Reeves,
%``On the Origin of the light Elements ($Z < 6$),"
%Rev. Mod. Phys., {\bf 66}, 193 (1994).
Reeves, H. 1994,
Rev. Mod. Phys., 66, 193 

\bibitem{reames67}                         % Needs of Data for Astrophysics
%Donald V. Reames,
%``Nuclear Cross Sections Required in Studies of Cosmic Rays,"
%in: {\em High-Energy Nuclear Reactions in Astrophysics},
%ed. B. Shen, New York: W. A. Benjamin (1967), pp. 273-281.
Reames, D. V. 1967,
in: High-Energy Nuclear Reactions in Astrophysics,
ed. B. Shen (New York: W. A. Benjamin), 273

\bibitem{ru66}
%G.~RUDSTAM,
%``Systematics of Spallation Yields,"
%{\em Z.~Naturforsch.~A}, {\bf 21}, 1027 (1966).
% No.~7, p.~1027--1041.
Rudstam, G. 1966,
Zs. f.~Naturforsch.~A, 21, 1027

\bibitem{ryan99}                   
% S. G. Ryan, T. C. Beers, K. A. Olive, B. D. Fields, and J. E. Norris,
%``Primordial Lithium and Big Bang Nucleosynthesis," 
% Preprint UMN-TH-1762/99; TPI-MINN-99/25; Eprint astro-ph/9905211.
Ryan, S. G., Beers, T. C., Olive, K. A., Fields, B. D., \& Norris, J. E. 1999,
preprint (astro-ph/9905211)

\bibitem{salpeter99}
%Edwin E. Salpeter,
%``Stellar Nucleosynthesis,"
%Rev. Mod. Phys., {\bf 71}, S220 (1999).
Salpeter, E. E. 1999,
Rev. Mod. Phys., 71, S220

\bibitem{sarkar99}        % The last, good review, on BBN (12 reactions !!!)
% Subir Sarkar,
% ``The Big Bang Nucleosynthesis: Reprise,"
% Eprint astro-th/9903183.
Sarkar, S.
1999, preprint (astro-th/9903183)

\bibitem{schramm95}
% David N. Schramm,
%``Primordial Nucleosynthesis and Light Element Abundances,"
%Proc. of an ESO/EIPC Workshop Held in Marciana Marina, Isola d'Ebla, 
%May 21--26, 1994,
Schramm, D. N. 1995,
in The Light Element Abundances,
Proc. of an ESO/EIPC Workshop Held in Marciana Marina, Isola d'Ebla, 
May 21--26, 1994,
ed. P. Crane (Berlin: Springer), 50

\bibitem{schramm98}
% David N. Schramm and Michael S. Turner, 
%C. Snede, R. N. Boyd, B. S. Meyer, and D. L. Lambert,
%``Big-Bang Nucleosynthesis Enters the Precision Era,"
%Rev. Mod. Phys., {\bf 70}, No. 1 (1998) 303-318 (1997).
Schramm, D. N., \& Turner M. S. 1998,
Rev. Mod. Phys., 70, 303 

\bibitem{newyield}
%R. Silberberg, C. H. Tsao, and A. F. Barghouty,
%``Updated Partial Cross Sections of Proton-Nucleus Reactions,"
%Astrophys. J., {\bf 501} (1998) 911-919. 
Silberberg, R., Tsao, C. H., \& Barghouty, A. F. 1998,
ApJ, 501, 911

% \bibitem{yield} %    
%R. Silberberg and  C. H. Tsao, 
%``Partial Cross-Sections in High-Energy Nuclear Reactions, and
%Astrophysical Applications. I. Targets With Z $\le 28$,"
%Astrophys. J. Suppl., No. 220, {\bf 25} (1973) 315-333;  
%R. Silberberg and  C. H. Tsao, 
%``Partial Cross-Sections in High-Energy Nuclear Reactions, and
%Astrophysical Applications. II. Targets Heavier Than Nickel,"
%Astrophys. J. Suppl., No. 220, {\bf 25} (1973) 335-368;  
%C. H. Tsao, private communication (1997).
\bibitem{yield73a}     
Silberberg, R., \& Tsao, C. H. 1973a,
ApJS, 220, 315

\bibitem{yield73b} 
Silberberg, R., \& Tsao, C. H. 1973b,
ApJS, 220, 335

\bibitem{yield76} 
Silberberg, R., Tsao, C. H., \& Shapiro, M. M. 1976,
in Astrophysics and Space Science Library, Vol. 59,
eds. B. S. O. Shen, \& M. Merker
(Dordrecht: D. Reidel Publ. Comp.), 49

\bibitem{simpson83}
% J. A. Simpson,
% ``Elemental and Isotopic Composition of the Galactic Cosmic Rays,"
% Ann. Rev. Nucl. Part. Sci., {\bf 33}, 323 (1983).
Simpson, J. A. 1983, 
Ann. Rev. Nucl. Part. Sci., 33, 323 

\bibitem{sisterson97}
%J. M. Sisterson et al., 
%K. Kim, A. Beverding, P. A. J. Englert, M. Caffee, A. J. T. Jull, 
%D. J. Donahue, L. McHargue, J. Vincent, and R. C. Reedy,
%       "Measurement of Proton Production Cross Sections of 10-Be and 26-Al
%       from Elements Found in Lunar Rocks,"
%{\em Nucl. Instr. Meth.}, {\bf B123} (1997) 324.
Sisterson, J. M.,  et al. 1997,
Nucl. Instr. Meth. B, 123, 324

\bibitem{smith96}
%Michael S. Smith (ORNL).
%F. Edward Cecil (Colorado School of Mines),
%Richard B. Firestone (LBNL),
%Gerald M. Hale (LANL),
%Duane C. Larson (ORNL),
%David A. Resler (LLNL),
%``U.S. Nuclear Data Resources for a Coordinated U.S. Effort in Nuclear Data
%for Nuclear Astrophysics,"
%http://www.dne.bnl.gov/~burrows/usnrdn/astrodata.html.
Smith, M. S., Cecil, F. E., Firestone, R. B., Hale, G. M., 
Larson, D. C., \& Resler, D. A. 1996,
U.S. Nuclear Data Resources for a Coordinated U.S. Effort in Nuclear Data
for Nuclear Astrophysics,
http://www.dne.bnl.gov/~burrows/usnrdn/astrodata.html

\bibitem{smith93}
%M. S. Smith, L. H. Kawano, and R. A. Malaney,
%``Experimental, Computational, and Observational Analysis of Primordial 
%Nucleosynthesis,"
%Astrophys. J. Suppl., {\bf 85} (1993) 219-247.  
Smith, M. S., Kawano, L. H., \& Malaney, R. A. 1993,
ApJS, 85, 219  

\bibitem{nuclex91} 
Sobolevsky, N. M. et al., 1991,
Production of Radionuclides at Intermediate Energies,
Subvolume A: Interaction of Protons with Targets from He to Br,
Landolt-B\"ornstein, New Series, I/13a,
ed. H. Schopper,
(Berlin, Heidelberg: Springer Verlag)

\bibitem{nuclex92} 
Sobolevsky, N. M. et al., 1992,
Production of Radionuclides at Intermediate Energies,
Subvolume B: Interaction of Protons with Targets from Kr to Te,
Landolt-B\"ornstein, New Series, I/13b,
ed. H. Schopper,
(Berlin, Heidelberg: Springer Verlag)

\bibitem{nuclex93} 
Sobolevsky, N. M. et al., 1993,
Production of Radionuclides at Intermediate Energies,
Subvolume C: Interaction of Protons with Targets from I to Am,
Landolt-B\"ornstein, New Series, I/13c,
ed. H. Schopper,
(Berlin, Heidelberg: Springer Verlag)

\bibitem{nuclex94a} 
Sobolevsky, N. M. et al., 1994a,
Production of Radionuclides at Intermediate Energies,
Subvolume D: Interaction of Protons with Nuclei
(Supplement to I/13a, b, c),
Landolt-B\"ornstein, New Series, I/13d,
ed. H. Schopper,
(Berlin, Heidelberg: Springer Verlag)

\bibitem{nuclex94b} 
Sobolevsky, N. M. et al., 1994b,
Production of Radionuclides at Intermediate Energies,
Subvolume E: Interaction of Pions and Antiprotons with Nuclei,
Landolt-B\"ornstein, New Series, I/13e,
ed. H. Schopper,
(Berlin, Heidelberg: Springer Verlag)

\bibitem{nuclex95} 
Sobolevsky, N. M. et al., 1995,
Production of Radionuclides at Intermediate Energies,
Subvolume F: Interaction of Deuterons, Tritons and 3He-nuclei with Nuclei,
Landolt-B\"ornstein, New Series, I/13f,
ed. H. Schopper,
(Berlin, Heidelberg: Springer Verlag)

\bibitem{nuclex96a} 
Sobolevsky, N. M. et al., 1996a,
Production of Radionuclides at Intermediate Energies,
Subvolume G: Interaction of $\alpha$-Particles with Targets from He to Rb,
Landolt-B\"ornstein, New Series, I/13g,
ed. H. Schopper,
(Berlin, Heidelberg: Springer Verlag)

\bibitem{nuclex96b} 
Sobolevsky, N. M. et al., 1996b,
Production of Radionuclides at Intermediate Energies,
Subvolume H: Interaction of $\alpha$-Particles with Targets from Sr to Cf,
Landolt-B\"ornstein, New Series, I/13h,
ed. H. Schopper,
(Berlin, Heidelberg: Springer Verlag)

\bibitem{NUCLEX_CD-ROM} 
Sobolevsky, N. M., et al., 2000,
Production of Radionuclides at Intermediate Energies,
Subvolume I: Interaction of Protons, Deuterons, Tritons, 3He-nuclei,
and $\alpha$-particles with Nuclei,
(Suppliment to volumes I/13 a-h),
Landolt-B\"ornstein, New Series, I/13i,
ed. H. Schopper,
(Berlin, Heidelberg: Springer Verlag)

\bibitem{steigman98}                       % A recent good review, on BBN 
% Gary Steigman,
% ``Big Bang Nucleosynthesis: Current Status"
% Eprint astro-th/9803055.
Steigman, G. 1998,
preprint (astro-th/9803055)

\bibitem{stepanov88}
%N.~V.~STEPANOV, 
%``Statistical Simulation of High-Excited Nuclei Fission.
%II.~Calculation and Comparison with Experiment," (in Russian)
%ITEP Preprint ITEP-55, Moscow (1988).
Stepanov, N. V. 1988,
ITEP Preprint ITEP-55, Moscow

\bibitem{eaf97}
Sublet, J. -Ch., Kopecky, J., Forrest, R. A., \& Niegro, D. 1997,
The European Activation File: EAF-97 Report file-Rev. 1,
UKAEA, Culham, Abigdon, Oxfordshire OX 14 3DB, United Kingdom

\bibitem{suess56}
%H. E. Suess and H. C. Urey,
%``Abundances of the Elements,"
%Rev. Mod. Phys., {\bf 28}, 53 (1956).
Suess, H. E., \& Urey, H. C. 1956,
Rev. Mod. Phys., 28, 53 

\bibitem{sudarqaim94}
Sudar, S. \& Qaim, S. M. 1994,
Phys. Rev. C, 50, 2408

\bibitem{summerer_blank00}
S\"ummerer, K. \& Blank, B. 2000,
Phys. Rev. C, 61, 034607

\bibitem{tassangot98}
%L. Tassan-Got et al., 
%B. Mustapha, F. Farget, M. Bernas, C. Stephan, P. Armbruster, J. Benlliure, 
%T. Enqvist, K. H. Schmidt, A. Boudard, R. Legrain, S. Leray, C. Volant, 
%W. Wlazlo, S. Czajkowski, and M. Pravikoff, 
%"Spallation Residue Cross-Sections in Reverse  Kinematics," 
%Proc. Int. Conf. on the Phys. of Nucl. Sci. and Techn.,
% October 5-8, 1998, Long Island, Ney York, vol. 2, pp. 1334-1340.
Tassan-Got, L.,  et al. 1998
in Proc. Int. Conf. on the Phys. of Nucl. Sci. and Techn.,
October 5-8, 1998, Long Island, Ney York, vol. 2
(LaGrange Park, IL: ANS Inc.), 1334

\bibitem{thomas93}
% David Thomas, David N. Schramm, Keith A. Olive, and Brian D. Fields,
%``Primordial Nucleosynthesis and the Abundances of Berillium and Boron,"
% Astrophys. Journ., {\bf 406} (1993) 569 (TSOF);
% Eprint arXiv:astro-ph/9209002.
Thomas, D., Schramm, D. N., Olive, K. A., \&Fields, B. D.
1993, ApJ, 406, 579

\bibitem{titarenko99a} 
% An example of how YIELD and several codes works: p(70 MeV)+Co-59 (co1.eps)  
%Yury E. Titarenko, Oleg V. Shvedov, 
%Vyacheslav F. Batyaev, Eugeny I. Karpikhin, Valery M. Zhivun,
%Ruslan D. Mulambetov, Andrey N. Sosnin,
%Stepan G. Mashnik, Richard E. Prael, 
%Tony A. Gabriel, and Marshall Blann,
%``Experimental and Theoretical Study of the Yields of Radionuclides
%Produced in $^{59}$Co Thin and Stacked Targets Irradiated 
%by 70-200 MeV Protons",
%Proc. 3rd Int. Conf. on Accelerator-Driven Transmutation Technologies
%and Applications (ADTTA'99), Praha, Czech Republic, June 7-11, 1999
%(Paper \# P-C27 on the Web page http://www.fjfi.cvut.cz/con_adtt99/). 
%LANL Report LA-UR-99-????; Eprint {\bf nucl-ex/9910???} 
%on LANL xxx.lanl.gov server.
Titarenko, Yu. E. et al. 1999a, in
Proc. 3rd Int. Conf. on Accelerator-Driven Transmutation Technologies
and Applications (ADTTA'99), Praha, Czech Republic, June 7-11, 1999,
(Paper \# P-C27 on the Web page http://www.fjfi.cvut.cz/con\_adtt99/)  

\bibitem{titarenko99b} 
%Yu. E. Titarenko, O. V. Shvedov, 
%V. F. Batyaev, E. I. Karpikhin, V. M. Zhivun,
%R. D. Mulambetov, A. N. Sosnin,
%S. G. Mashnik, R. E. Prael, 
%T. A. Gabriel, and M. Blann,
%``Experimental and Computer Simulation Study of Radionuclide
%Production in Heavy Materials Irradiated 
%by Intermediate Energy Protons",
%Proc. 3rd Int. Topical Meeting on Nuclear 
%Applications of Accelerator Technology (AccApp'99), 
%Long Beach, CA, November 14-18, 1999, 
%ANS, Inc., LaGrange Park, IL, pp. 212-221.
Titarenko, Yu. E. et al. 1999b,
in Proc. 3rd Int. Topical Meeting on Nuclear 
Applications of Accelerator Technology (AccApp'99), 
Long Beach, CA, November 14-18, 1999, 
(LaGrange Park, IL: ANS, Inc.), 212

\bibitem{tobailem71} 
Tobailem, J., Lassus St-Genies, C. H., \& Leveque, L. 1971,
Note CEA-N-1466(1), Saclay

\bibitem{tobailem75} 
Tobailem, J., \& Lassus St-Genies, C. H. 1975,
Note CEA-N-1466(3), Saclay

\bibitem{tobailem77} 
Tobailem, J., \& Lassus St-Genies, C. H. 1977,
Note CEA-N-1466(4), Saclay

\bibitem{tobailem81a} 
Tobailem, J. 1981a,
Note CEA-N-1466(5), Saclay

\bibitem{tobailem81b} 
Tobailem, J. 1981b,
Note CEA-N-1466(6), Saclay

\bibitem{tobailem82} 
Tobailem, J. 1982,
Note CEA-N-1466(7), Saclay

\bibitem{tobailem83} 
Tobailem, J. 1983,
Note CEA-N-1466(8), Saclay

\bibitem{tsao99}
%C. H. Tsao, A. F. Barghouty, and R. Silberberg,
%``Nuclear Cross Sections and the Composition, Transport, and Origin of
%Galactic Cosmic Rays,"
%in {\em Topics in Cosmic-Ray Astrophysics}, Horizonts in World Physics
%series, vol. 230, Nova Science Publishers, Inc., Commack, New York,
%1999, pp. 141-168.
Tsao, C. H., Barghouty, A. F., \& Silberberg R. 1999,
in Horizonts in World Physics Series, vol 230, 
Topics in Cosmic-Ray Astrophysics, 
(Commack, New York: Nova Science Publishers, Inc. ), 141

\bibitem{tsipenyuk97}
% Yuri M. Tsipenyuk,
% ``Nuclear Methods in Science and Technology,"
%{\em Fundamental and Applied Nuclear Physics Series},
% Edited by David A. Bradley,
%Institute of Physics Publishing, Bristol (1997).
Tsipenyuk, Yu. M. 1997,
Nuclear Methods in Science and Technology,
Fundamental and Applied Nuclear Physics Series,
ed. D. A. Bradley,
(Bristol, UK: Institute of Physics Publishing)

\bibitem{turner96}
Turner, M. S. et al. 1996,
Astrophys. J. Lett., 466, L59

\bibitem{turner99}
%Michael S. Turner,
%``Cosmological Parameters,"
Turner, M. S. 1999,
in The Proc. of Particle Physicsand the Universe (Cosmo-98),
edited By O. Caldwell (AIP, Woodbury, NY), to be published

\bibitem{turnertyson99}
%Michael S. Turner and J. Anthony Tyson,
%``Cosmology at the Millenium,"
%Rev. Mod. Phys., {\bf 71}, S145 (1999).
Turner, M. S. \& Tyson, J. A. 1999,
Rev. Mod. Phys., 71, S145

\bibitem{tytlet00}
% David Tytler, John M. O'Meara, Nao Suzuki, and Dan Lubin,
%``Review of Big Bang Nucleosynthesis and Primordial Abundances,"
% Eprint: arXiv:astro-ph/0001318; to be published in Physica Scripta.
Tytler, D., O'Meara, J. M., Suzuki, N., \& Lubin, D. 2000,
Physica Scripta, in press, preprint (arXiv:astro-ph/0001318)

\bibitem{VangioniFlam96}
% Elisabet Vangioni-Flam, Michel Cass\'e, Brian D. Fields, and Keith A. Olive,
%``LiBeB Production by Nuclei and Neutrinos,"
% Astrophys. Journ., {\bf 468} (1996) 199-206.
Vangioni-Flam, E., Cass\'e, M., Fields, B. D., \& Olive, K. A.
1996, ApJ, 468, 199

\bibitem{VangioniFlam99}
% Elisabet Vangioni-Flam, Michel Cass\'e, and Jean Audouze,
% ``Lithium-Berilium-Boron: Origin and Evolution,"
% Eprint astro-th/9907171.
Vangioni-Flam, E., Cass\'e, M., \& Audouze, J.
1999, preprint (astro-th/9907171)

\bibitem{VangioniFlam00}
% Elisabet Vangioni-Flam, Alain Coc, Michel Cass\'e, and Yvette Oberto,
% ``Big Bang Nucleosynthesis updated with NACRE Compilation,"
% Eprint arXiv:astro-th/0002248.
Vangioni-Flam, E., Coc, A., Cass\'e, M., \& Oberto, Y. 2000,
preprint (arXix:astro-th/0002248)

\bibitem{vanriper97}
Van Riper, K. A., et al. 1997,
LANL Report LA-UR-97-5068

\bibitem{vanriper98}
Van Riper, K. A., Mashnik, S. G., \& Wison, W. B. 1998,
LANL Report LA-UR-98-5379 (a 684 page detailed report with 37
tables and 264 color figures is available at the Web page
http://t2.lanl.gov/publications/publication.html)

\bibitem{vanriper00}
Van Riper, K. A., Mashnik, S. G., \& Wison, W. B. 2000,
Nucl. Instr. Meth. A, in press (nucl-th/9901073)

\bibitem{venikov93}
% N. I. Venikov, V. I. Novikov, and A. A. Sebiakin, 
% "Excitation Functions
%      of Proton-induced Reactions on 126-Xe: 125-I Impurity in 123-I,"
%{\em Appl. Radiat. Isot.}, {\bf 44} (1993) 751. 
Venikov, N. I., Novikov, V. I., \& Sebiakin, A. A. 1993,
Appl. Radiat. Isot., 44, 751 

\bibitem{vonach97}
%H. Vonach et al., 
%A. Pavlik, A. Wallner, M. Drosg, R. C. Haight, D. M. Drake, and S. Chiba, 
%"Spallation Reactions in Al-27 and Fe-56 Induced by 800 MeV Protons," 
%{\em Phys. Rev.}, {\bf C55} (1997) 2458.
Vonach, H.,  et al. 1997,
Phys. Rev. C, 55, 2458

\bibitem{waddington99}
%C. Jake Waddington,
%``Source Abundances and Nuclear Cross Sections,
%in {\em Topics in Cosmic-Ray Astrophysics}, Horizonts in World Physics
%series, vol. 230, Nova Science Publishers, Inc., Commack, New York,
%1999, pp. 199-212.
Waddington, C. J. 1999,
in Horizonts in World Physics Series, vol 230, 
Topics in Cosmic-Ray Astrophysics, 
(Nova Science Publishers, Inc., Commack, New York), 199

%\bibitem[]{} 
\bibitem{40years} 
%G. Wallerstein, I. Parker, Jr., A. M. Boesgaard, G. M. Hale, A. E. Champagne,
%C. A. Barnes, F. K\"{a}ppeler, V. V. Smith, R. D. Hoffman, F. X. Timmes,
%C. Snede, R. N. Boyd, B. S. Meyer, and D. L. Lambert,
%``Synthesis of the Elements in Stars: Forthy Years of Progress,"
%Rev. Mod. Phys., {\bf 69}, 995 (1997).
Wallerstein, E. et al. 1997,
Rev. Mod. Phys., 69, 995 

\bibitem{webber90}
%W. R. Webber, J. C. Kish, and D. A. Schrier, 
%"Individual Isotopic
%       Fragmentation Cross Sections of Relativistic Nuclei in Hydrogen,
%       Helium, and Carbon Targets," 
%{\em Phys. Rev.}, {\bf C41} (1990) 547.
Webber, W. R., Kish, J. C., \& Schrier, D. A. 1990,
Phys. Rev. C, 41, 547

\bibitem{weigel99}
%Andreas Weigel, Otto Eugster, Christian Koeberl, Rolf Michel, Urs Krahenbuhl,
%and Sonja Neumann,
%``Relationship Amoung Lodranites and Acapulcoites: Noble Gas Isotope
%Abundances, Chemical Composition, Cosmic-Ray Exposure ages, and Solar
%Cosmic Ray Effects,"
%Geochimica et Cosmochimica Acta, Vol. 62, No. 2, pp. 175-192, 1999.
Weigel, A. et al. 1999,
Geochimica et Cosmochimica Acta, 62, 175

\bibitem{wlazlo00}
%W. Wlazlo, T. Enqvist, P. Armbruster, J. Benlliure, M. Bernas, A. Boudard, 
%S. Czajkowski, R. Legrain, S. Leray, B. Mustapha, M. Pravikoff, F. Rejmund, 
%K.-H. Schmidt, C. Stephan, J. Taieb, L. Tassan-Got, and C. Volant,
%``Cross-Sections of Spallation Residues Produced in 1$\cdot$QA GeV
%$^{208}$Pb on Proton Reactions,"
%Eprint: arXiv:nucl-ex/0002011 (2000), submited to Phys. Rev. Lett.  
Wlazlo, W., et al. 2000,
Phys. Rev. Lett., in press (arXiv:nucl-ex/0002011)

\bibitem{wolfenstain99}
% L. Wolfenstein,
%``Neutrino Physics,"
%Rev. Mod. Phys., {\bf 71}, No. 2, S140-S144 (1999).
Wolfenstein, L. 1999,
Rev. Mod. Phys., 71, S140

\bibitem{woosley90}
% S. E. Woosley, D. H. Hartmann, R. D. Hoffman, and W. C. Haxton,
%``The $\nu$-Process,"
% Astrophys. Journ., {\bf 356} (1990) 272-301.
Woosley, S. E., Hartmann, D. H., Hoffman, R. D., \& Haxton, W. C.
1990, ApJ, 356, 272

\bibitem{woosley95}
% S. E. Woosley and Thomas A. Weaver,
% ``The Evolution and Explosion of Massive Stars. II. Explosive Hydrodynamics
% and Nucleosynthesis,"
% The Astrophysical Journal Supplement Series, {\bf 101} (1995) 181-235.
Woosley, S. E., \& Weaver T. A.
1995, ApJS, 101, 181

\end{thebibliography}
\end{document}